\documentclass{lmcs}  
\pdfoutput=1

\usepackage{lastpage}
\lmcsdoi{20}{1}{22}
\lmcsheading{}{\pageref{LastPage}}{}{}%
{Jun.~16,~2023}{Mar.~20,~2024}{}


\usepackage{hyperref}
 
\theoremstyle{plain}  

\usepackage[utf8]{inputenc}
\usepackage{microtype}
\usepackage{booktabs}
\usepackage{tikz}
\usepackage{xspace}
\usepackage[algoruled]{algorithm2e}
\usetikzlibrary{shapes,arrows,positioning}

\usepackage[shortlabels]{enumitem}

\usepackage{amstext}
\usepackage{amsmath}
\usepackage{amsfonts}

\usepackage{amssymb}

\usepackage{colortbl}
\usepackage{slashbox}

\usepackage{graphicx}
\usepackage{subfig}

\newcommand{\nat}{\mathbb{N}}
\newcommand{\integer}{\mathbb{Z}}
\newcommand{\rat}{\mathbb{Q}}
\newcommand{\Lan}{\mathcal{L}}

\newcommand{\cA}{\mathcal{A}}
\newcommand{\cM}{\mathcal{M}}
\newcommand{\cN}{\mathcal{N}}

\definecolor{verylow}{HTML}{e75874}
\definecolor{low}{HTML}{e75874}
\definecolor{medium}{HTML}{ff8c00}
\definecolor{high}{HTML}{53A567}
\definecolor{veryhigh}{HTML}{53A567}

\newcolumntype{"}{@{\hskip\tabcolsep\vrule width 0.8pt\hskip\tabcolsep}}

\newcommand{\N}{\mathbb{N}}
\newcommand{\A}{\mathcal{A}}

\newcommand{\emptyword}{\lambda}
\renewcommand{\epsilon}{\varepsilon}

\begin{document}

\title[Analyzing Robustness of Angluin's L$\!^*$ Algorithm in Presence of Noise]{Analyzing Robustness of Angluin's\texorpdfstring{\\}{ }L$\!^*$ Algorithm in Presence of Noise}
 
\author[L.~Ye et al]{Lina Ye\lmcsorcid{0000-0002-2217-4752}}[a]
\author[]{Igor Khmelnitsky\lmcsorcid{0000-0002-5132-5839}}
\author[]{Serge Haddad\lmcsorcid{0000-0002-1759-1201}}[b]
\author[]{Benoît Barbot\lmcsorcid{0000-0003-2417-3064}}[c]
\author[]{Benedikt Bollig\lmcsorcid{0000-0003-0985-6115}}[d]
\author[]{Martin Leucker\lmcsorcid{0000-0002-3696-9222}}[e]
\author[]{Daniel Neider\lmcsorcid{0000-0001-9276-6342}}[f,g]
\author[]{Rajarshi Roy\lmcsorcid{0000-0002-0202-1169}}[h]

\address{Université Paris-Saclay, CNRS, ENS Paris-Saclay, CentraleSupélec, LMF, France} 
\email{lina.ye@lmf.cnrs.fr}

\address{Université Paris-Saclay, CNRS, ENS Paris-Saclay, LMF, France} 
\email{serge.haddad@lmf.cnrs.fr}  

\address{Université Paris-Est Créteil, France}	 

\address{Université Paris-Saclay, CNRS, ENS Paris-Saclay, LMF, France}	  

\address{Institute for Software Engineering and Programming Languages, Universität zu
Lübeck, Germany}

\address{ 
 TU Dortmund University, Germany}
 
 \address{ 
Center for Trustworthy Data Science and Security, University Alliance 
Ruhr, Germany}

\address{Max Planck Institute for Software Systems, Kaiserslautern, Germany}

\begin{abstract}
  \noindent    Angluin's L$^\ast$ algorithm learns the minimal
          deterministic finite automaton (DFA) of a regular language
          using membership and equivalence queries. Its probabilistic
          approximatively correct (PAC) version substitutes an
          equivalence query by numerous random membership
          queries to get a high level confidence to the answer. Thus
          it can be applied to any kind of device
          and may be viewed as an algorithm for synthesizing an
          automaton abstracting the behavior of the device based on
          observations.  Here we are interested on how Angluin's PAC
          learning algorithm behaves for devices which are obtained
          from a DFA by introducing some noise. More precisely we
          study whether Angluin's algorithm reduces the noise and
          produces a DFA closer to the original one than the noisy
          device.  We propose several ways to introduce the noise: (1)
          the noisy device inverts the classification of words w.r.t.\
          the DFA with a small probability,  (2) the noisy
          device modifies with a small probability the letters of the
          word before asking its classification w.r.t.\ the DFA, (3) the
          noisy device combines the classification of a word w.r.t.\ the DFA
          and its classification w.r.t. a counter automaton, and (4) the noisy DFA is obtained by a random process from two DFA such that the language of the first one is included in the second one. Then when a word is accepted (resp. rejected) by the first (resp. second) one, it is also accepted (resp. rejected) and in the remaining cases, it is accepted with probability 0.5.   
          Our main experimental contributions consist in showing that: (1)
          Angluin's algorithm behaves well whenever the noisy device 
         is produced by a random process, (2) but poorly with a structured noise, and, that (3) is able to eliminate pathological behaviours specified in a regular way.
         Theoretically, we show that randomness almost surely yields systems with non-recursively enumerable languages.
\end{abstract}

\maketitle

\section*{Introduction}
\label{sec:intro}

\paragraph{Discrete-event systems and their languages.}

Discrete-event systems \cite{Cassandras10} form a large class of dynamic systems that, given some internal state, evolve from one state to another one due
to the occurrence of an event. For instance, discrete-event systems can represent a  cyber-physical process whose events are triggered 
by a controller or the environment, or, a business process whose events are triggered by human activities or software executions.
Often, the behaviors of such systems are classified as safe (aka correct, representative, etc.) or unsafe. Since a behavior may be identified
by its sequence of occurred events, this leads to the notion of a language.

\paragraph{Analysis versus synthesis.}

There are numerous formalisms to specify (languages of) discrete-event systems.
From a designer's perpective, the simpler it is the better its analysis will be. So finite automata and their languages (regular languages)
are good candidates for the specification thanks to their simplicity. However, even when the system is specified by an automaton, its implementation may slightly differ due to several reasons (bugs, unplanned human activities, unpredictable environment, etc.). Thus, one generally checks whether 
the implementation conforms to the specification. However, in many contexts, the system under consideration has already been implemented and the original specification
(if any) is lost, as for instance in the framework of process mining \cite{vanderalst12}. Thus, by observing and interacting with the system, one aims to recover a specification close to the system at hand but that is robust with respect to its pathologic behaviors.   

\paragraph{Language learning.}

The problem of learning a language from finite samples of strings by discovering the corresponding grammar is known as grammatical inference. Its significance was initially stated in~\cite{Solomonoff64} and an overview of very first results can be found in~\cite{Biermann72}. As it may not always be possible to infer a grammar that exactly identifies a language, approximate language learning was introduced in~\cite{Wharton1974}, where a grammar is selected from a solution space whose language approximates the target language with a specified degree of accuracy.  To provide a deeper insight into language learning, the problem of identifying a (minimal) deterministic finite automaton   (DFA) that is consistent with a given sample has attracted substantial attention in the literature since several decades~\cite{Biermann72b, Gold1978, Angluin87, Valiant84,Natasha20,Furelos21}.  An understanding of regular language learning is very valuable for a generalization to other more complex classes of languages. For example, Some researchers adapted learning algorithms for regular languages to learn context-free languages~\cite{Clark07,Clark2010,Yoshinaka2010}.

\paragraph{Angluin's L$\!^{*}$ algorithm.}

Angluin's L$\!^{*}$ algorithm learns the minimal DFA of a regular language with two types of queries: membership queries and equivalence queries~\cite{Angluin87}. Angluin's approach triggered a lot of subsequent research on active automata learning and has numerous applications, such as finding bugs in implementations of security-critical protocols~\cite{Brostean16,Brostean17,Brostean17b}, learning interfaces of classes in software libraries~\cite{HowarISBJ12}, and inferring interface protocols of legacy software components~\cite{Kousar20}. One could of course try to adapt it to the synthesis task described above. Since 1987, different improvements of the original Angluin's $L^*$ algorithm have been proposed, thus resulting in numerous variants~\cite{Rivest89,Kearns94,Maler91,Shahbaz09,Bollig09}. It is fair to say that L$\!^{*}$-like algorithms completely dominate the research area of active automata learning. 
However, for most black box systems, it is often impossible to implement the equivalence query.  Thus, its probabilistic approximatively correct (PAC) version substitutes an equivalence query with a large enough set of random membership queries. Using a PAC framework, one needs to define and evaluate the accuracy of such an approach. Hence, here we are interested in how PAC Angluin's algorithm behaves for devices which are obtained from a DFA by introducing some noise.

\paragraph{Noisy learning.}

Most learning algorithms in the literature assume the correctness of the training data, including the example data such as attributes as well as classification results. However, sometimes noise-free datasets are not available. \cite{Quinlan86} carried out an experimental study of the noise effects on the learning performance. The results showed that generally the classification noise had more negative impact than the attribute one, i.e., errors in the values of attributes.     
\cite{AngluinL87} studied how to compensate for randomly introduced noise and discovered a theorem giving a bound on the  smaple size that is sufficient  for PAC-identification in the presence of classification noise when the concept classes are finite.  
Michael Kearns formalized another related learning model from statistical queries by extending Valiant's learning model~\cite{Kearns98}. One main result shows that any class of functions learnable from this statistical query model is also learnable with classification noise in Valiant's model. 

\paragraph{Our contribution.}

\begin{figure}[h]
	\centering
	\includegraphics[scale=0.9]{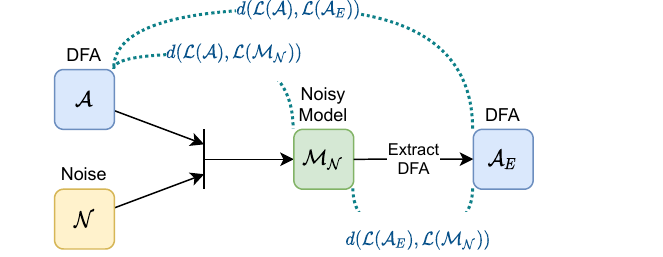}
	
	\caption{The experimental setup and the studied distances}
	
	\label{fig:General}
\end{figure}

In this paper, we study against which kinds of noise Angluin's algorithm is \emph{robust}. In this work we use the optimized version of  this algorithm from \cite{Kearns94}. So to avoid confusion, we will call it the KV's algorithm.  To the best of our knowledge, this is the very first attempt of noise analysis in the automata learning setting.
More precisely, we consider the following setting (cf.\ Figure~\ref{fig:General}): Assume that a regular device $\cA$ is given, typically as a black box. Due to some noise $\cN$, the system $\cA$ is pertubed resulting in a not necessarily regular system $\cM_\cN$. This one is consulted by the PAC version of KV's algorithm to obtain a regular system $\cA_E$. The question studied in this paper is whether $\cA_E$ is closer to $\cA$ than $\cM_\cN$, or, in other words, to which extent learning via KV's algorithm  is robust against the noise $\cN$.
 
To this end, we introduce four kinds of \emph{noisy devices} obtained from $\A$: 
\begin{enumerate}
    \item the noisy device is obtained by a random process from a given DFA by inverting the classification of words with a small probability, which corresponds to the classification noise in the classical learning setting;  
    \item the noisy device is obtained by a random process that, with a small probability, replaces each letter of a word by one chosen uniformly from the alphabet and then determines its classification based on the DFA, which corresponds to the attribute noise in the classical setting; 
    \item the noisy  DFA combines the classification of a word w.r.t. the DFA and its status w.r.t.\ a counter automaton;   
    \item The noisy DFA is obtained by a random process from two DFA such that the language of the first one is included in the second one. Then when a word is accepted (resp. rejected) by the first (resp. second) one, it is also accepted (resp. rejected) and in the remaining cases, it is accepted with probability 0.5.
\end{enumerate}

Our studies are based on the distribution over words that is used for generating words associated with membership queries 
and defining (and statistically measuring) the \emph{distance}
between two devices as the probability that they differ on word acceptance. 
We have performed experiments
over several hundreds random DFA. 
We have pursued several goals along our experiments, expressed by the following questions:
 
\begin{itemize}[]
\setlength\itemsep{0.5em}
	\item What is the threshold (in terms of distance) between pertubating the DFA or producing a device that is no more ``similar to'' the DFA?
	\item What is the impact of the nature of noise on the robustness of  KV's algorithm?
	\item What is the impact of the words distribution on the robustness of  KV's algorithm?
        \item How to reduce the size of the extracted DFA by  KV's algorithm while keeping as close as possible to the original DFA?
	
\end{itemize}
 
Due to the approximating nature of the PAC version of $L^\ast$, we had to consider the question of how to choose the accuracy  of the approximate equivalence query to get a good trade-off between accuracy and efficiency. Moreover, since in most cases,  KV's algorithm may perform a huge number of refinement rounds before a possible termination, we considered what a ``good'' number of rounds to stop the algorithm avoiding underfitting and overfitting is.

We experimentally show that w.r.t. the random noise, i.e., the noise is randomly introduced, KV's algorithm behaves quite well, i.e., the learned DFA ($\cA_E$) is very often closer to the original one ($\cA$) than the noisy random device ($\cM_\cN$). When the noise is obtained using the counter automaton, KV's algorithm is not robust. Instead, the device $\cA_E$ is closer to the noisy device $\cM_\cN$. Moreover, we establish that the expectation of the length of a random word should be large enough to cover a relevant part of the set of words in order for KV's algorithms to be robust. The size of the extracted DFA can be further reduced by returning an intermediate memorized DFA which is enough close to the  DFA returned by the KV's algorithm. 

In order to understand why  KV's algorithm is robust w.r.t. random noise we have undertaken a theoretical study establishing that
almost surely the language of the noisy device  ($\cM_\cN$) for classification noise, i.e., case (1) and, with a further weak assumption on DFA, also for instance noise, i.e.,  case (2) is not recursively enumerable. Considering non-recursively enumerable languages as unstructured, this means that due to the noise, the (regular) structure of $\cA$ vanishes. This is not the case for the counter automaton setting. Altogether, to put it bluntly: the less structure the noisy device has, the better KV's algorithm works.

 \paragraph{Organization.}

In Section~\ref{sec:preliminaries}, we introduce the technical background required for the robustness analysis.
In Section~\ref{sec:algorithm}, we detail the goals and the settings of our analysis. In Section~\ref{sec:evaluation}, we provide and discuss the experimental results. 
In Section~\ref{sec:theory}, we discuss randomness versus structure. 
Finally in Section~\ref{sec:con}, we draw our the conclusions and identify future work.

\section{Preliminaries} 
\label{sec:preliminaries}
 
Here we provide the technical background required for the robustness analysis.

\paragraph{Languages.}

Let $\Sigma$ be an alphabet, i.e., a nonempty finite set,
whose elements are called \emph{letters}. A \emph{word} $w$ over $\Sigma$
is a finite sequence over $\Sigma$, whose length is denoted by $|w|$.
The unique word of length $0$ is called the \emph{empty word} and denoted by $\emptyword$.
As usual, $\Sigma^\ast$ is the set of all words over $\Sigma$,
and $\Sigma^+ = \Sigma^\ast \setminus \{\emptyword\}$ is
the set of words of positive length.
A \emph{language} (over $\Sigma$)
is any set $L \subseteq \Sigma^\ast$.
The symmetric difference of languages $L_1,L_2 \subseteq \Sigma^\ast$
is defined as $L_1 \Delta L_2 = (L_1 \setminus L_2) \cup (L_2 \setminus L_1)$.

\paragraph{Words distribution and measure of a language.}

A distribution $D$ over $\Sigma^*$ is defined by a mapping ${\bf Pr}_D$
from $\Sigma^*$ to $[0,1]$ such that $\sum_{w\in \Sigma^*}{\bf Pr}_D(w)=1$.
Let $L$ be a language. Its \emph{probabilistic measure} w.r.t. $D$, ${\bf Pr}_D(L)$ 
is defined by ${\bf Pr}_D(L)=\sum_{w\in L}{\bf Pr}_D(w)$, i.e., the sum of the probabilities 
for all words in $L$.

Our analysis requires that we are able to efficiently  sample a word according to 
some distribution $D$. Thus we only consider  distributions $D_\mu$ with  $\mu\in\ ]0,1[$,
that are defined for a word $w = a_1 \ldots a_n \in \Sigma^\ast$ by
\[{\bf Pr}_{D_\mu}(w) = \mu \left(\frac{1-\mu}{|\Sigma|}\right)^n\,.\]

To sample a random word according to $D_\mu$ in practice, we start with the empty word
and iteratively we flip a biased coin with probability $1-\mu$ to add a letter (and $\mu$ to return the current word)
and then uniformly select the letter in $\Sigma$.

\paragraph{Language distance.}

Given two languages $L_1$ and $L_2$, their distance w.r.t.\ a distribution $D$,
$d_D(L_1,L_2)$, is defined by $d_D(L_1,L_2)={\bf Pr}_D(L_1 \Delta L_2)$, the sum of the probabilities for 
all words that are only in $L_1$ or only in $L_2$. Computing the distance between
languages is in most of the cases impossible. Fortunately whenever the membership problem
for $L_1$ and $L_2$ is decidable, then using Chernoff-Hoeffding bounds~\cite{Hoff63}, this distance can be statistically approximated as follows.
Let $\alpha,\gamma > 0$ be an error parameter and a confidence level, respectively.
Let $S$ be a set of words sampled independently according to $D$, called a sampling, such that  $|S| \geq \frac{\log(2 / \gamma )}{ 2\alpha^2}$.
Let
$dist =\frac{|S \cap (L_1 \Delta L_2)|}{|S|}$.
Then, we have
\[
{\bf Pr}_D(|d_D(L_1,L_2) - dist| > \alpha ) ~<~ \gamma\,.
\]
Since we will not simultaneously discuss about multiple distributions,
we omit the subscript $D$
almost everywhere.

\paragraph{Finite Automata.}

A (complete) deterministic finite automaton (DFA) over $\Sigma$ is a tuple
$\A = (Q,\sigma,q_0,F)$ where $Q$ is a finite set
of states, $q_0 \in Q$ is the initial state,
$F \subseteq Q$ is the set of final states, and
$\sigma: Q \times \Sigma \to Q$ is the transition function.
The transition function is inductively extended over words by 
 $\sigma(q,\emptyword)=q$ and  $\sigma(q,wa)=\sigma(\sigma(q,w),a)$.
The language of $\A$ is defined as
$\Lan(\A) = \{w \in \Sigma^\ast \mid \sigma(q_0,w) \in F\}$.
A language $L \subseteq \Sigma^\ast$ is called \emph{regular} if
$L = \Lan(\A)$ for some DFA $\A$.

\newcommand{\tprob}{p}
\newcommand{\ntprob}{p}
\newcommand{\letterprop}{p}
\newcommand{\width}{\epsilon}
\newcommand{\confidence}{\gamma}
\newcommand{\letterprobp}[1]{p_{#1}}

\paragraph{KV's algorithm.}

Given a regular language $L$, KV's  algorithm learns the unique mimimal DFA $\A$ such that $\Lan(\A)=L$ through two queries:
\begin{itemize}
\item membership queries: `Does a word $w$ belong to $L$?' 
\item equivalence queries: `Does $\Lan(\A_E)=L$? and if not provide a word $w\in L\Delta \Lan(\A_E)$'. Here $\A_E$ represents an automaton synthetised by the algorithm based on currently collected membership queries.  
\end{itemize}
The goal of this algorithm is to produce a DFA that recognizes a given regular 
language $L\subseteq \Sigma^*$. It is worth noting that while $\Sigma$ is given, 
$L$ is a priori unknown and can only be accessed through membership queries and 
equivalence queries. 
Let us describe in an informal way its behaviour:
\begin{enumerate}
\item  Initialize a data structure $Data$ to store useful information from an initial set of membership queries.
\item Repeat the following steps:
\begin{itemize}
\item synthetize a DFA based on the current $Data$ corresponding to the set of membership queries currently asked (see \cite{Kearns94} for more details);
\item submit an equivalence query, for which the algorithm can return a pair consisting of a boolean value and a string as a potential counter example: either {\bf true} with empty string or 
{\bf false} with a counter example $w$;
\item when not equivalent and using the counter example $w \in L\Delta \Lan(\A_E)$, update $Data$.
\end{itemize}
\end{enumerate}
The KV's algorithm ensures that after each round
the number of states of the current DFA is increased by one.
So the number of rounds is equal to the number of states of the minimal DFA accepting $L$.

\paragraph{PAC version of KV's  algorithm.}

The KV's algorithm is efficient to learn the minimal DFA when both queries are available and there is no noise in the data. 
However, as our goal is to analyze the robustness of this algorithm against noise, the latter should be introduced into our data, which results in some noisy 
device that is not necessarily regular. To handle this, we resort to Probably Approximately Correct (PAC) learning framework, whose goal 
is to guarantee with high probability that the hypothesis will have low generalization error. Precisely, the learner can learn 
the concept, i.e., here DFA, given any arbitrary approximation ratio and success probability. 
The PAC model was later extended to treat noise~\cite{Sloan95,Decatur97,BSHOUTY02}.

The PAC version of KV's algorithm takes as input an error parameter $\varepsilon$ and a confidence level $\delta$, and 
replaces the equivalence query by a (large enough) number of membership queries `$w \in  L\Delta \Lan(\A)$?', where the words are sampled from some distribution $D$ unknown to the algorithm. Thus this algorithm can stop too early when all answers are negative while $L\neq \Lan(\A)$. However due to the large   number of such queries which depends on the current round $r$ (i.e., $\lceil \frac{log(1/\delta)+(r+1)\log(2)}{\varepsilon}\rceil$) this algorithm ensures that
\[
{\bf Pr}_D(d_D(L,\Lan(\A)) > \epsilon ) ~<~ \delta\,.
\]

A key observation is that the PAC version of this algorithm could be used for every language $L$ for which the membership problem is decidable. However, in some cases, 
$L$ is not necessarily a regular language, the algorithm might never stop and thus our adaptation includes a parameter $maxround$, the upper bound on the number of rounds, that ensures termination.
This is formalized by Algorithm~\ref{algo:PAC}. Observe that if either the automaton is wrongly considered as equivalent or the maximal round is reached, then $\A_E$ is not necessarily minimal and
so the algorithm minimizes $\A_E$ before returning it.

\begin{algorithm}[h]

        \DontPrintSemicolon
        
        \KwIn{$L$, a language unknown to the algorithm}
        \KwIn{an integer $maxround$ ensuring termination}
        \SetKwFunction{KV}{KV}
        \SetKwFunction{Minimize}{Minimize}

	\BlankLine
	\KwData{an integer $r$, a boolean $b$, a data structure $Data$ and a DFA $\A_E$}
	\KwOut{a DFA}
	\BlankLine
	
	{\tt Initialize}($Data$)\;
	$r \leftarrow 0$\;
	\BlankLine
	
	\tcp{The control of $maxround$ is unnecessary when $L$ is regular}
	\While{$r < maxround$}
	{
	 $\A_E \leftarrow $ {\tt Synthetize}($Data$)\;
	 
	 $(b,w) \leftarrow$ {\tt IsEquivalent}($\A_E$)
	  
	\lIf{$b$}
	{\Return {\Minimize}($\A_E$)}
	 {\tt Update}($Data,w$)\;
	 $r\leftarrow r+1$
	}
	
	 \Return  {\Minimize}({\tt Synthesize}($Data$))

	\caption{The PAC version of KV's algorithm applied on general languages}
	\label{algo:PAC}    
\end{algorithm}

\section{Robustness Analysis against Noises}
\label{sec:algorithm}
\subsection{Principle and goals of the robustness analysis}
\label{subsec:goals}

\paragraph{Principle of the analysis.}
Figure~\ref{fig:General} illustrates the process of our robustness analysis with respect to different types of noises that will be studied in this paper. 
First we set the qualitative and quantitative nature of the noise ($\mathcal N$). 
Then we generate a set of random DFA ($\A$).
Combining $\A$ and $\mathcal N$, one gets a noisy model $\mathcal M_{\mathcal N}$.
More precisely, depending on whether the noise is random or not,  
$\mathcal M_{\mathcal N}$ is either generated off-line (deterministic noise) or on-line
(random noise) when a membership query is asked during the execution of KV's algorithm. 
Finally we compare (1) the distances between $\A$ and $\mathcal M_{\mathcal N}$, 
and (2) between $\A$ and $\mathcal  \A_E$, the automaton returned by
the algorithm. The aim of this comparison is to establish whether
$\A_E$ is closer to $\A$ than $\mathcal M_{\mathcal N}$. 
In order to get a quantitative measure, we define the  \emph{information gain} as: 
\[
	\text{Information gain} = \frac{d(\Lan(\A),\Lan(\mathcal M_{\mathcal N}))}{d(\Lan(\A),\Lan(\A_E))}
\] 
 
We consider a {\bf \color{low} low} information gain to be in $[0
,0.9)$, a {\bf\color{medium} medium} information gain to be in $[0.9
,1.5)$, and a  {\bf\color{high} high} information gain to be in $[1.5 ,\infty)$. 
After preliminary experiments, these thresholds seem
to appropriately partition the results into three subsets of relevant sizes. 
The higher the information gain is, the closer $\A_E$ is to $\A$ than $\mathcal M_{\mathcal N}$.

In addition,
we also evaluate the distance between  $\A_E$ and $\mathcal M_{\mathcal N}$
in order to study in which cases the algorithm learns in fact the noisy device instead of the original DFA.

\paragraph{Goals of the analysis.}

\begin{itemize}[] \setlength\itemsep{0.7em}
	\item {\bf Quantitative analysis.} The information gain highly depends on the `quantity' of the noise, which is also called error rate. So we analyze the information gain 
	depending of the distance between the original DFA and the noisy device and want to identify
	a threshold (if any) where  the information gain starts to significantly increase.
	\item {\bf Qualitative analysis.} Another important criterion of the information gain is the `nature' of the noise. So we analyze the information gain w.r.t. the different noisy devices that we have introduced.
	\item{\bf Impact of word distribution.} Then, the
          robustness of the KV's algorithm with respect to word distribution is also analyzed.
\end{itemize}

 In order to perform relevant experiments, one needs to tune two critical parameters of KV's algorithm. Since  the running time of the algorithm quadratically depends on
	the number of rounds (i.e. iterations of the loop), selecting an appropriate \emph{maximal number of rounds} is a critical issue.
	We vary this maximal number of rounds and analyze how the information gain decreases
	w.r.t. this number.
On the other hand, as an equivalence query is replaced with a set of membership queries whose number depends
	on the current round and the pair $(\varepsilon,\delta)$, it is thus interesting to study (1) what is the
	effect of \emph{accuracy of the approximate equivalence queries}, i.e., the values of  $(\varepsilon,\delta)$ on the ratio of executions that reach the maximal number of rounds
	and (2) compare the  information gain  for executions that stop before reaching this maximal number
	and the same execution when letting it run up  to this maximal number.

\subsection{Noise}

\newcommand{\RL}{R}

A \emph{random language} $R \subseteq \Sigma^\ast$ is determined by
a random process: for each $w \in \Sigma^\ast$, membership $w \in \RL$
is determined independently at random, \emph{once and for all}, according to
some probability ${\bf Pr}(w \in R) \in [0, 1]$. The probability ${\bf Pr}(w \in R)$
may depend on some parameters such as $w$ itself and a given DFA.

We now describe the  four types of noise that we analyze in this paper.
Each type adds noise to a given DFA $\A$ in form of a random language $\RL$.
For the first two types, \emph{noise with output} and \emph{noise with input},
the probability ${\bf Pr}(w \in R)$ of including $w \in \Sigma^\ast$ in $\RL$ depends on $w$
itself, $\Lan(\A)$, and some parameter $0<p<1$.
The third kind of noise, \emph{counter DFA}, is actually \emph{deterministic},
i.e., ${\bf Pr}(w \in R) \in \{0,1\}$ for all $w \in \Sigma^\ast$. In that case,
the given DFA $\A$ determines a unique ``noisy'' language.
Let us be more precise:

\paragraph{DFA with noisy output.}

Given a DFA $\A$ over the alphabet $\Sigma$ and $0<p<1$, the random
language $\Lan(\A^{\rightarrow p})$ flips the classification of words w.r.t.\ $\Lan(\A)$ with probability $p$.
More formally, for all $w \in \Sigma^\ast$, $${\bf Pr}(w\in \Lan(\A^{\rightarrow p}))=(1-p){\bf 1}_{w\in\Lan(\A)}+p{\bf 1}_{w \not\in\Lan(\A)}$$ 
where ${\bf 1}_{C}$ is 1 if condition $C$ holds, and 0 otherwise.
Observe that the expected value of the distance  
$d(\Lan(\A),\Lan(\A^{\rightarrow p}))$ is $p$. Moreover, in our experiments, 
we observe that $\left|\frac{d(\Lan(\A),\Lan(\A^{\rightarrow p}))-p}{p}\right|<5\cdot 10^{-2}$ for all the generated languages.

\paragraph{DFA with noisy input.}

Given a DFA $\A$ over the alphabet $\Sigma$ (with $|\Sigma|>1$) and $0<p<1$, the random language
$\Lan(\A^{\leftarrow p})$ 
changes every letter of the word with probability $p$
uniformly to another letter and then returns the classification of the new word w.r.t.\ $\Lan(\A)$. More formally, for $w=a_1\ldots a_n \in \Sigma^\ast$,
$${\bf Pr}(w\in \Lan(\A^{\leftarrow p}))=\sum_{\substack{w'=b_1\ldots b_n\in\Lan(\A) \\\textup{s.t. }|w|=|w'|}}~\prod_{1 \le i\leq n} \Bigl((1-p){\bf 1}_{a_i=b_i}+\frac{p}{|\Sigma|-1} {\bf 1}_{a_i\neq b_i}\Bigr)\,.$$

\paragraph{Counter DFA}
Let $\A$  be a DFA over the alphabet $\Sigma$ and $c: \Sigma \cup \{\lambda\} \to \integer$ be a function.
We inductively define the function $\overline{c}: \Sigma^* \to \integer$ by
$$ 
	\overline{c}(\lambda)= c(\lambda) \mbox{ and } \overline{c}(wa)= \overline{c}(w)+c(a)\,.
$$
The counter language $\Lan(\A_c)$ is now given as the union of the language of $\A$ and the set of words whose value of function $\overline{c}$ is nonpositive. 
$$\Lan(\A_c) = \Lan(\A) \cup \{w \in \Sigma^\ast \mid \overline{c}(w)\leq 0\}\,.$$

\paragraph{DFA with pathological behaviours} The forth device that we want to learn   
is a DFA $\A$ viewed as a formal model for a protocol that should
be followed by users in some institution (hospital, university, etc.). 
We only consider  $\A$ for which there is a word denoted $w_\A$
such that $w_\A\Sigma^* \cap \Lan(\A)=\emptyset$.
This is a reasonable assumption for a realistic protocol
for which after a specific \emph{pathological} sequence of actions (here $w_{\A}$), 
one knows that the user cannot succeed.

The observed language generated by the noisy (random) device, denoted here $\A^n$, 
is obtained as follows:
\begin{itemize}
    \item $\Lan(\A^n)$ includes $\Lan(\A)$. The observed language
    must contain all correct behaviours w.r.t. the protocol;
    \item For every word $w\in  w_\A\Sigma^*$,
    $w \in \Lan(\A^n)$ with probability $\frac 1 2$. Every pathological 
    behaviour may equally be observed or not;
    \item For every word $w\notin \Lan(\A) \uplus w_\A\Sigma^*$, $w \notin \Lan(\A^n)$. One does not observe any behaviour which is neither correct nor pathological. 
\end{itemize}

\section{Experimental Evaluation}
\label{sec:evaluation}

In order to empirically evaluate our ideas, we have implemented a prototype and benchmarks in Python, using the NumPy library. 
They are available on Zenodo\footnote{\url{https://doi.org/10.5281/zenodo.8031255}}.
All evaluations were performed on a computer equipped by Intel i5-8250U CPU with 4 cores, 16GB of memory and Ubuntu Linux~18.03.

\subsection{Generating DFAs}
\label{sec:generateDFA}

We now describe the settings of the experiments we made with  four different types of noises. 
We choose $\mu=10^{-2}$ for the parameter of the word distribution so that the average length
of a random word is $99$.
All the statistic distances were computed using the Chernoff-Hoeffding bound~\cite{Hoff63} 
with $\alpha = 5\cdot 10^{-4}$ as error parameter and $\gamma = 10^{-3}$ as confidence level. 

The benchmarks were performed on  DFA randomly generated using the following procedure.
Let $M_q=60$ and $M_a=20$ be two parameters, which impose upper bounds on the number of states and the size of the alphabet, that could be tuned in future experiments. 
The DFA $\A=(Q,\sigma,q_0,F)$ on $\Sigma$ is generated as follows:
\begin{itemize} \setlength\itemsep{0.5em}
	\item Uniformly choose $n_q\in[20,M_q]$ and $n_a\in[3,M_a]$;
	\item Set $Q = [0,n_q]$ and $\Sigma = [0,n_a]$;
	\item Uniformly choose $n_f\in[0,n_q-1]$ and let $F = [0,n_f]$;
	\item Uniformly choose $q_0$ in $Q$;
	\item For all $(q,a)\in Q\times\Sigma$, choose the target state $\sigma(q,a)$ uniformly among all states.	
\end{itemize}
The choice of $M_q$ and $M_a$ was inspired by observing that these values often occur when
 modeling  realistic processes like in business process 
management.

\subsection{Tunings} \label{subsec:tunings}

Before launching our experiments, we first tune two key parameters for both efficiency and accuracy purposes:
the maximal number of rounds of the algorithm and  the value of $ \epsilon $ and of $ \delta $ for the accuracy of the approximate equivalence query.
For the sake of simplicity, this tuning is based on experiments over the DFA with the noisy output. The reason is that the expected distance between the DFA and the noisy device is known and can be controlled ($p$).

\paragraph{Maximal number of rounds.}

\begin{figure}[h]
	\centering
 	 \subfloat{{\includegraphics[width=7.1cm]{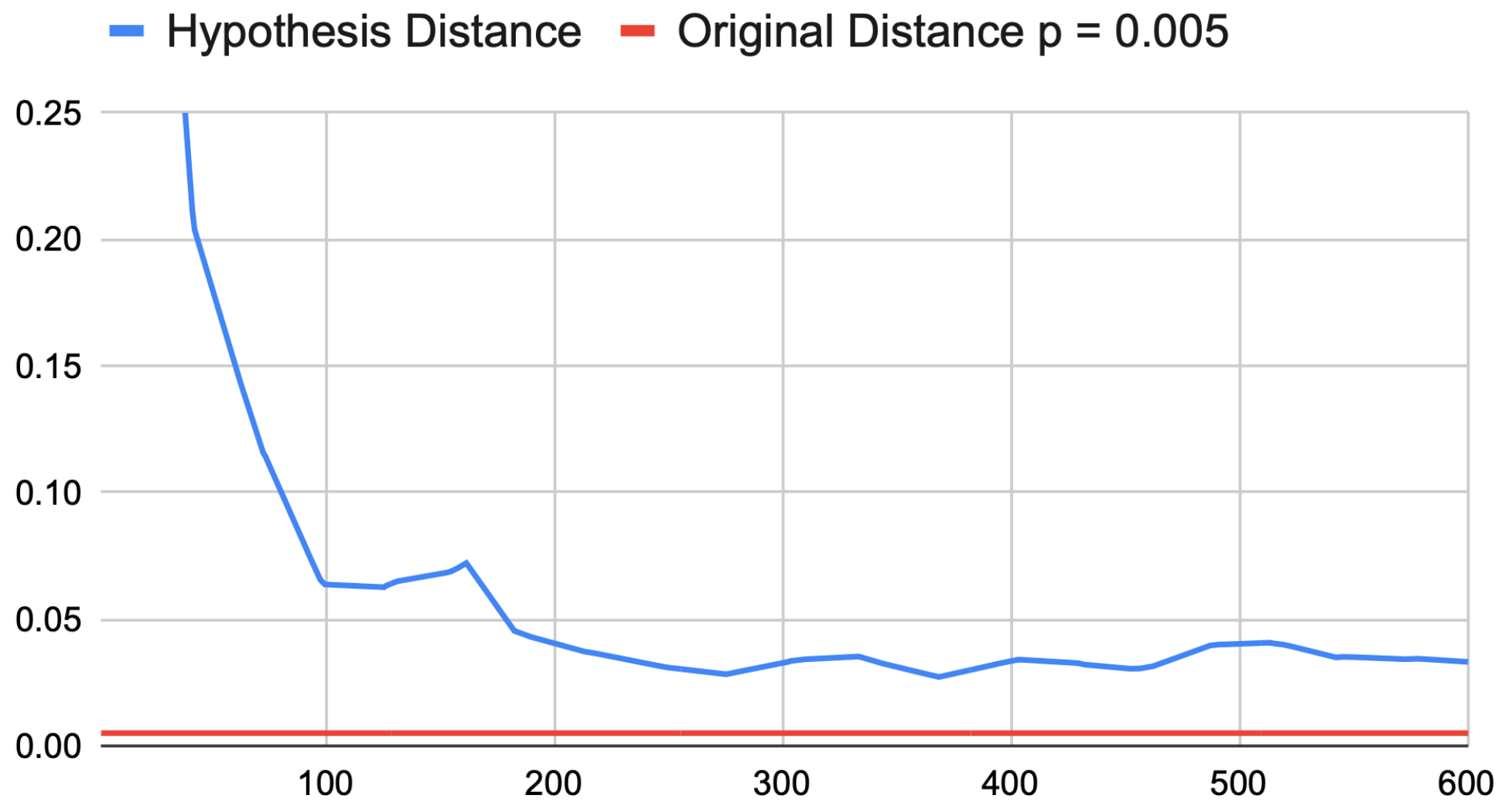} }}
    \qquad
    \subfloat{{\includegraphics[width=7.1cm]{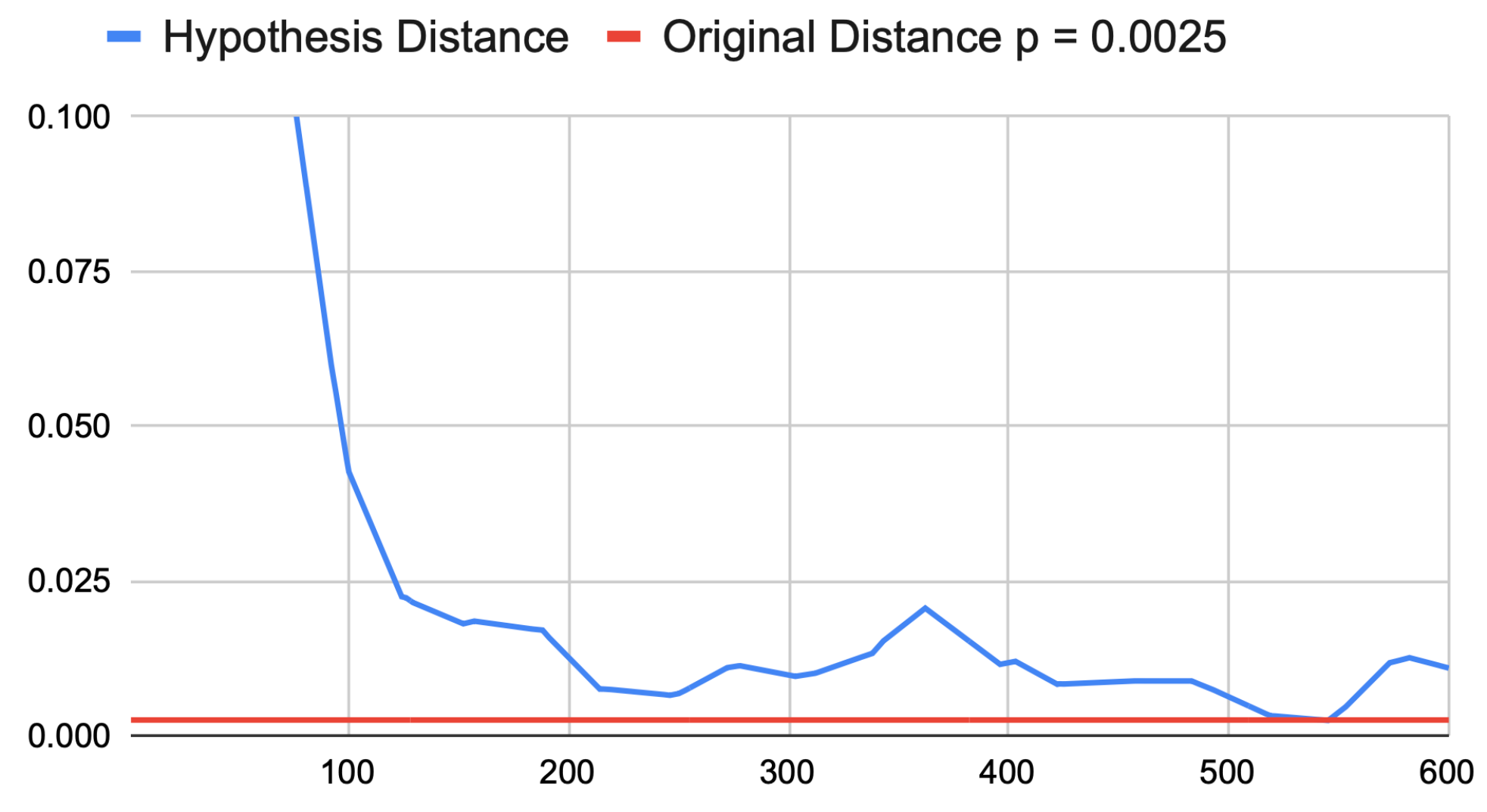} }}
 \qquad
 
 	 \subfloat{{\includegraphics[width=7.1cm]{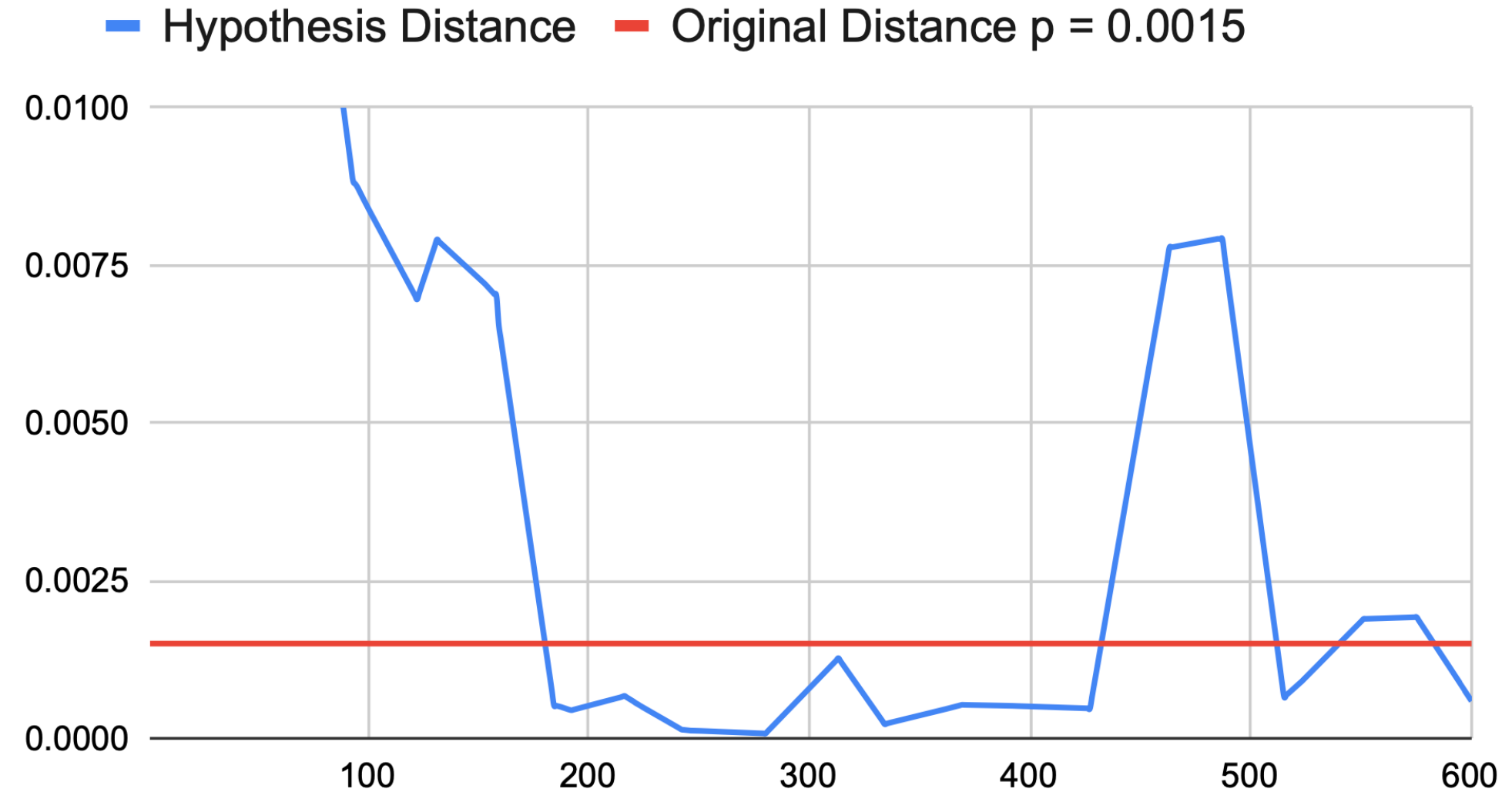} }}
    \qquad
    \subfloat{{\includegraphics[width=7.1cm]{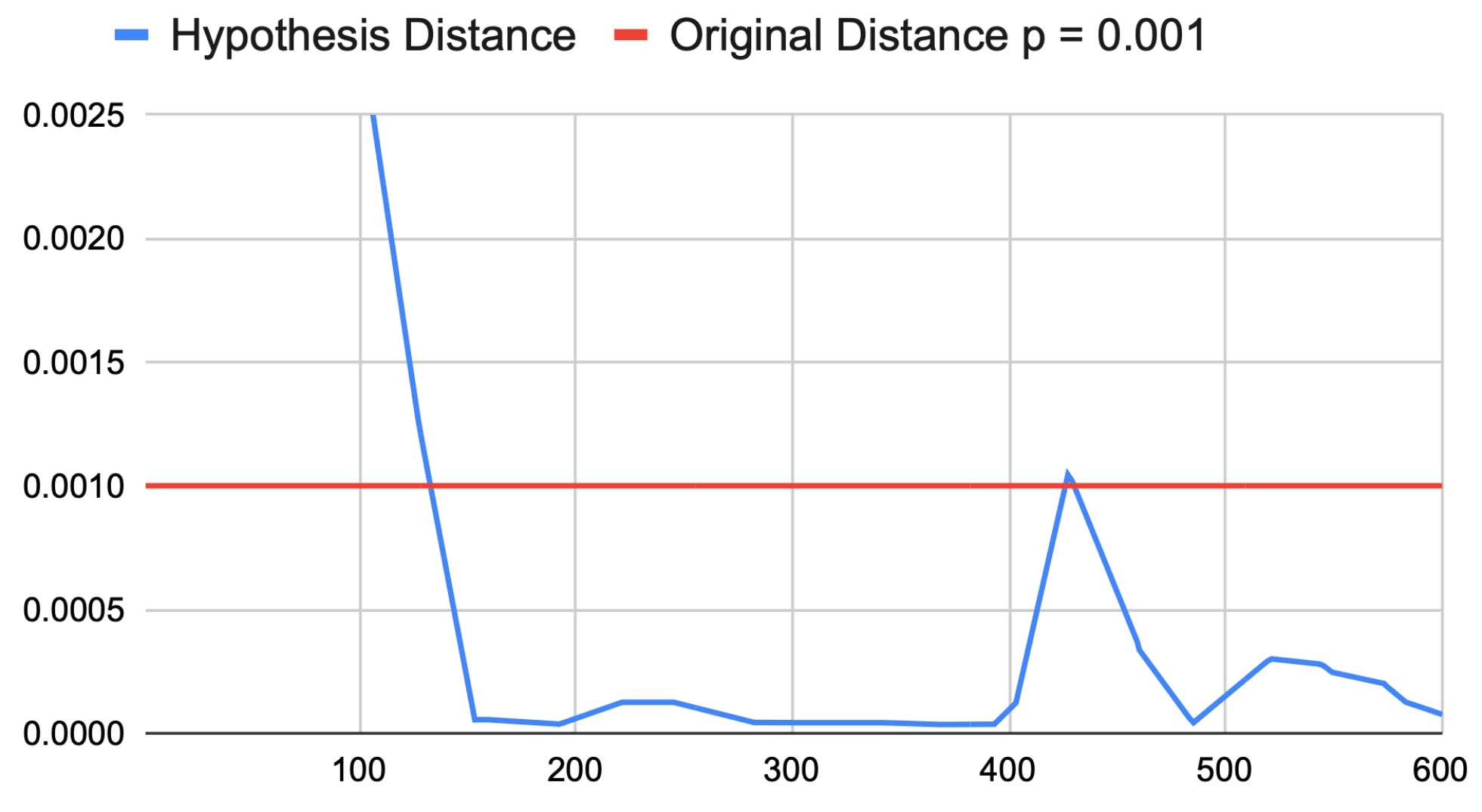} }}
	\caption{Number of rounds analysis}
	\label{fig:roundAnalysis1}
\end{figure}

In order to specify a maximal number of rounds that lead to a good performance of the KV's Algorithm, we took a DFA with noisy output 
$\A^{\rightarrow p}$ for $p\in\{0.005,0.0025,0.0015,0.001\}$.
We ran the learning algorithm, stopping every 20 rounds to estimate the distance between the current DFA $\A_E$ to the original DFA $\A$.
Figure~\ref{fig:roundAnalysis1} shows the evolution graphs of  $d(\Lan(\A),\Lan(\A_E))$ w.r.t. the number of rounds 
according to the different values of $p$, each of them summarizing five runs on five different DFAs. 
The vertical axis corresponds to the distance to original DFA $\A$, and the horizontal axis corresponds to  the number of rounds. 
The red line is the distance with $\A^{\rightarrow p}$, and the blue line is the distance with $\A_E$.

We observe that after about 250 rounds, the distance $d(\Lan(\A),\Lan(\A_E))$ is stabilizing except some rare peaks, which are worth further investigation. 
Therefore, from now on all the experiments are made with a maximum of 250 rounds.
Of course this number depends on the size of $\A$. However, for the variable size that we have chosen (between 10 and 50 states), 
it seems to be a good choice. 

\paragraph{Accuracy of the approximate equivalence query.}
We have generated thirty-five DFA, and for each of them we generated five $\A^{\rightarrow p}$ with different values of $p$.
Table~\ref{tbl:epsilonAndDelatEpirementet} summarizes our results with different $\epsilon$ and $\delta$ for the approximate equivalence query. 
The rows correspond to the value of the noise $p$, the columns correspond to the values of $ \epsilon $ and $ \delta $ (where we always choose $\epsilon=\delta$) and each cell shows the average information gain. From this table, 
$\varepsilon=\delta=0.01$ and $\varepsilon=\delta=0.005$ seem to be optimal values. We decided to fix  $\varepsilon=\delta=0.005$
for all our experiments.
 
	\begin{table}[h]
		\centering
		\scalebox{.9}{
		{\renewcommand{\arraystretch}{1.5}
		\begin{tabular}{l*{5}{@{\hskip1em}r}}
			\toprule
			\backslashbox{$p$}{$\varepsilon=\delta$} & $0.05$ & $0.01$ & $0.005$ & $0.001$ & $0.0005$
			\tabularnewline
			\midrule
			$0.01$ &\cellcolor{low}$0.081$ & \cellcolor{low}$0.054$ & \cellcolor{low}$0.047$ & \cellcolor{low}$0.048$ & \cellcolor{low}$0.050$ \tabularnewline
			$0.005$ & \cellcolor{low}$0.086$ & \cellcolor{low}$0.087$ & \cellcolor{low}$0.072$ & \cellcolor{low}$0.070$ & \cellcolor{low}$0.094$ \tabularnewline
			$0.0025$ & \cellcolor{low}$0.867$ & \cellcolor{low}$0.292$ & \cellcolor{low}$0.591$ &\cellcolor{low} $0.321$ & \cellcolor{low}$0.748$ \tabularnewline
			$0.0015$ & \cellcolor{medium}$1.401$ & \cellcolor{high}$2.933$ & \cellcolor{high}$3.082$ & \cellcolor{medium}$0.980$ & \cellcolor{low}$0.710$ \tabularnewline
			$0.001$ & \cellcolor{high}$5.334$ & \cellcolor{high}$4.524$ & \cellcolor{high}$3.594$ & \cellcolor{high}$1.811$ & \cellcolor{high}$6.440$ \tabularnewline
			\bottomrule
		\end{tabular}}
		}
		\caption{ Evaluation of the impact of $\epsilon$ and $\delta$. }\label{tbl:epsilonAndDelatEpirementet}
	\end{table}

\subsection{Qualitative and Quantitative analysis}
For the four types of noise, we have generated numerous DFA (as described shortly above). Furthermore, for the noisy outputs and inputs given a randomly generated  DFA, we have constructed several noisy devices depending on the `quantity' of noise.
By computing the (average) information gain for all these experiments, we have been able to get conclusions about the effect of the nature and 
the quantity of the noise on the performance of KV's algorithm.

When KV's algorithm is applied to a device with random noise, a corresponding
random language is generated on-the-fly: once membership of a word in the
target language has been determined (e.g., through a membership query),
the corresponding truth value is stored and not changed anymore.

\paragraph{DFA with noisy output.}
We have generated fifty DFA, and for each such DFA $\A$, we have generated random languages
with noisy output $\Lan(\A^{\rightarrow p})$ with five values for $p$ between $0.01$ and $0.001$.
Table~\ref{tbl:noisyoutput} summarizes the results.
Recall that the expected value of $d(\mathcal L(\A),\mathcal L(\A^{\rightarrow p}))$ is $p$.
We have identified a threshold for $p$ around  $0.0025$:
if the noise is higher than $0.0025$, the resulting DFA $\mathcal{A}_E$ has a bigger distance to the original one $\A$ than to the corresponding noisy device $\A^{\rightarrow p}$, and smaller otherwise.
Moreover, once we cross the threshold the robustness of the algorithm increases very quickly. We have also included a column
that represents the standard \textit{deviation} of the random variable $d(\mathcal L(\A),\mathcal L(\A_E))$ to assess that our conclusions are robust w.r.t.
the probabilistic feature.

	\begin{table}[h]
		\centering
		\label{tbl:summaryBenchmarkNoisy}
		\scalebox{.9}{
		{\renewcommand{\arraystretch}{1.5}
		\begin{tabular}{l*{4}{r}}
			\toprule
			$ p $ &$d(\mathcal L(\A),\mathcal L(\A_E))$ & $d(\mathcal L(\A^{\rightarrow p}),\mathcal L(\A_E))$ &gain  & deviation\tabularnewline
			\midrule
			$0.01$ & $0.12625$ & $0.13320$ &\cellcolor{low} $0.07432$ & $0.04102$	\tabularnewline
			$0.005$ & $0.04420$ & $0.04827$ & \cellcolor{low} $0.11312$  & $0.03366$	\tabularnewline
			$0.0025$ & $0.00333$ & $0.00568$ &\cellcolor{low} $0.75031$ & $0.00523$ \tabularnewline
			$0.0015$ & $0.00027$ & $0.00174$ &\cellcolor{high} $5.52999$ & $0.00047$ \tabularnewline
			$0.001$ &  $0.00006$ & $0.00103$ &\cellcolor{high} $15.75817$& $0.00007$ \tabularnewline
			\bottomrule
		\end{tabular}}
	}
		
		\caption{Evaluation of the algorithm w.r.t.\ the noisy output.}\label{tbl:noisyoutput}
	\end{table}

\paragraph{DFA with noisy input.}
We have generated forty-five random DFA, and for each such DFA $\A$,
we have generated random languages with noisy input $\Lan(\A^{\leftarrow p})$ by choosing $p\in\{10^{-4},5\cdot10^{-4},10^{-3},5\cdot10^{-3}\}$. 
Contrary to the case of noisy output, $p$ does not correspond to the expected value of $d(\mathcal L(\A),\mathcal L(\A^{\leftarrow p})$.
Thus we need to calculate this distance for every pair of the experiments. Thus, we have gathered the pairs whose distances belong 
to intervals that are described in the first column of Table~\ref{tbl:summaryBenchmarkNoisyInput}.
The second column of this table reports the number of pairs in the interval while the third one presents  
the average value of the distance for the set of the corresponding pairs.
Again we identify a threshold for $d(\mathcal L(\A),\mathcal L(\A^{\leftarrow p}))$ between $0.001$ and $0.005$. 
Once we cross the threshold, the robustness of the algorithm increases very quickly.

	\begin{table}[h]
		\centering

		\scalebox{.9}{
		{\renewcommand{\arraystretch}{1.5}
		\begin{tabular}{l*{6}{r}}
			\toprule
			Range & \# &$d(\mathcal L(\A),\mathcal L(\A^{\leftarrow p}))$  &$d(\mathcal L(\A),\mathcal L(\A_E))$ &$d(\mathcal L\A^{\leftarrow p}),\mathcal L(\A_E))$ &gain&  deviation\tabularnewline
			\midrule 
			$[0.025, 1]$ & $36$ & $0.04027$ & $0.21513$ & $0.22658$ & \cellcolor{low}$0.18$& 0.05279\tabularnewline
			$[0.005, 0.025]$ & $53$ & $0.00924$ & $0.05416$ & $0.06077$ &\cellcolor{low} $0.17$ & 0.04172\tabularnewline
			$[0.002,0.005]$ &33& $0.00378$ & $0.01260$ & $0.01611$ &\cellcolor{low} $0.30$ & 0.01783\tabularnewline
			$[0.001,0.002]$ & $11$ & $0.00123$ & $0.00030$ & $0.00154$ &\cellcolor{high} $4.1$ & 0.00058\tabularnewline
			$[0.0005,0.001]$ & $25$ & $0.00079$& $0.00002$ & $0.00082$&\cellcolor{high} $39.5$ & 0.00007\tabularnewline	
			\bottomrule
		\end{tabular}}
	}

		\caption{Evaluation of the algorithm w.r.t.\ the noisy input.}\label{tbl:summaryBenchmarkNoisyInput}
	\end{table}

\paragraph{Counter DFA}
We have randomly generated the counter function as follows:
We have uniformly chosen $c(\lambda)$ in $[0,|\Sigma|]$. Then,
for all $a \in \Sigma$, ${\bf Pr}(c(a)=-1)=\frac 1 4$ and for all $0\leq i \leq 6$, ${\bf Pr}(c(a)=i)=\frac 3 {28}$.

We have generated 160 DFA. For each of them, we have generated a counter automaton (as described before). 
The results of our experiments are given in Table~\ref{tbl:counterDFA}, from which we can see that whatever the quantity of noise, the KV's algorithm is unable to get closer to the original DFA. Moreover the extracted DFA $\A_E$ is very often closer to the counter automaton $\A_c$ than the original DFA $\A$. 
	
	\begin{table}[h]
		\centering
		\scalebox{.9}{
		{\renewcommand{\arraystretch}{1.5}
		\begin{tabular}{l*{6}{r}}
			\toprule
			Range& \# & $d(\mathcal L(\A),\mathcal L(\A_{c}))$  &$d(\mathcal L(\A),\mathcal L(\A_E))$ & 
			$d(\mathcal L(\A_c),\mathcal L(\A_E))$ &gain&  deviation \tabularnewline
			\midrule
			$[0.005, 0.025]$ & $14$ & $0.01238$ & $0.02586$ & $0.02053$ & \cellcolor{low}$0.47886$ & $0.01898$ \tabularnewline
			$[0.002, 0.005]$ & $57$ & $0.00245$ & $0.00396$ & $0.00262$ & \cellcolor{low}$0.61765$ & $0.00298$ \tabularnewline
			$[0.001, 0.002]$ & $22$ & $0.00143$ & $0.00209$ & $0.00121$ & \cellcolor{low}$0.68156$ & $0.00126$ \tabularnewline
			$[0.0005, 0.001]$ & $20$ & $0.00079$ & $0.00108$ & $0.00064$ & \cellcolor{low}$0.72481$ & $0.00065$ \tabularnewline
			$[0.0001, 0.0005]$ & $44$ & $0.00025$ & $0.00035$ & $0.00021$ & \cellcolor{low}$0.71054$ & $0.00021$ \tabularnewline
			\bottomrule
		\end{tabular}}
			}

		\caption{Evaluation of the algorithm w.r.t.\ the `noisy' counter.}\label{tbl:counterDFA}
	\end{table}

Thus we conjecture that when the noise is `unstructured' and the quantity is small enough such that the word noise
is still meaningful, then KV's algorithm is robust against such noise. On the contrary when the noise is structured, then KV's algorithm
`tries to learn' the noisy device whatever the quantity of noise. In Section~\ref{sec:theory}, we will strengthen this conjecture
establishing that in some sense, noise produced by random process implies unstructured noise.

\subsection{DFA with pathological behaviours}
\label{sec:superset}

Let $\A^+$ be a DFA such that $\Lan(\A^+)=\Lan(\A) \uplus w_\A\Sigma^*$.
Theoretically, $d(\Lan(\A^n),\Lan(\A^+))$ is on average equal to $d(\Lan(\A^n),\Lan(\A))$.
Thus a priori, the KV's algorithm applied on $\A^n$
should not modify this
property since $\A$ and $\A^+$ are DFA. 
However due to the particular shape of the pathological behaviours,
it could be not the case. So the goal of studying this device is to detect the general tendancy: is the output DFA closer to $\A$ or to $\A^+$?
In order to be relevant, 
${\bf Pr}_d(w_{\A}\Sigma^*)=\left(\frac{1-\mu}{|\Sigma|}\right)^{|w_{\A}|}$ should be sufficiently small.

\paragraph{DFA construction}
We now explain how to randomly generate a DFA $\A$
such that there exists a word $w_\A$ fulfilling $w_\A\Sigma^*\cap \Lan(\A)=\emptyset$. Since we do not care
about the letters of $w_\A$, we choose $w_\A=a^3$ and randomly
select $|\Sigma|$ with $5\leq |\Sigma| \leq 20$ which  
leads to ${\bf Pr}_d(w_{\A}\Sigma^*)\leq \frac 1 {125} \leq 0.008$. As for the other parameters, we adopt the same as described before, for instance the number of states is between 20 and 60.
\begin{enumerate}
    \item We first create states $q_0,q_1,q_2 \in Q$
    and $q_\bot\in Q\setminus F$
    with $\sigma(q_0,a)=q_1$, $\sigma(q_1,a)=q_2$, $\sigma(q_2,a)=q_\bot$,
    and for all $b\in \Sigma$, $\sigma(q_\bot,b)=q_\bot$;
    \item Then we proceed as in section~\ref{sec:generateDFA}
    for achieving the construction of the rest of $\A$ but excluding $\{q_0,q_1,q_2,q_\bot\}$
    to be the destinations of a transition.

\end{enumerate}
Such a DFA $\A$ is depicted in Figure~\ref{fig:A1} with $\Sigma=\{a, b\}$, for the sake of simplicity.
$\A^+$ is equal to $\A$ except that $q_\bot$ is now accepting.

\begin{figure}[h]
\begin{footnotesize}
\begin{center}
\begin{tikzpicture}[node distance=1.5cm and 1.5cm]
  \tikzset{
    sh2s/.style={shift={(0,-1)}},
    rc/.style={rounded corners=3mm,line width=1pt},
    place/.style={draw,circle,inner sep=2pt,minimum size=20pt, align=center},
  }
  \node (00) {};
  \node[place, node distance=1cm, right=of 00] (0) {$q_0$};
  \node[place, right=of 0] (1) {$q_1$};
  \node[place, right=of 1] (2) {$q_2$};
  \node[place, right=of 2] (3) {$q_\bot$};
  \node[place, node distance=1cm, double, below=of 1] (4) {$q_4$};
  \node[place, node distance=1cm, below=of 2] (5) {$q_5$};
  \node[place, node distance=1cm, double, right=of 5] (6) {$q_6$};
 \draw[arrows=-latex'] (00) -- node[above, align=center]{}  (0);
 \draw[arrows=-latex'] (0) -- node[above, align=center]{$a$}  (1);
 \draw[arrows=-latex'] (1) -- node[above, align=center]{$a$}  (2);
  \draw[arrows=-latex'] (2) -- node[above, align=center]{$a$}  (3);
 \draw[arrows=-latex'] (1) -- node[right, align=center]{$b$}  (4);
 \draw[arrows=-latex'] (2) -- node[right, align=center]{$b$}  (5);
 \draw [arrows=-latex'] (3) edge[loop above]node{$a, b$} (3);
 \draw[arrows=-latex'] (0) -- node[above right, align=center]{$b$}  (4);
 \draw[arrows=-latex'] (4) -- node[above, align=center]{$a$, $b$}  (5);
 \draw[arrows=-latex'] (5) to[bend right] node[above, align=center]{$b$}  (6);
 
 \draw [arrows=-latex'] (6) to[bend right] node[above, align=center]{$a$} (5);
  \draw [arrows=-latex'] (5) edge[loop below]node{$a$} (5);
  \draw [arrows=-latex'] (6) edge[loop right]node{$b$} (6);
   
\end{tikzpicture}
\end{center}
\end{footnotesize}
\caption{A DFA $\A$ where $a^3\Sigma^*\cap \Lan(\A)=\emptyset$.}
\label{fig:A1}
\end{figure}

\paragraph{Experimental results}
 
We have constructed and run about 300 benchmarks, each one being a pair of ($\A$,$\A^+$). 
 In this context, the information gain is defined as follows.    

\[
	\text{Information gain} = \frac{d(\Lan(\A^+),\Lan(\A^n))}{d(\Lan(\A),\Lan(\A^n))}
\]

The goal of our experiments is to check whether the gain is strictly bigger than 1 or not. Different from Section~\ref{subsec:goals}, we only consider here  a {\bf\color{low} low} information gain when it is less than one
and a {\bf\color{high} high} one when the information gain otherwise.

	\begin{table}[h]
		\centering

		\scalebox{.9}{
		{\renewcommand{\arraystretch}{1.5}
		\hspace*{-7pt}
		\begin{tabular}{l*{6}{r}}
			\toprule
			Range & \# &$d(\mathcal L(\A),\mathcal L(\A^+))$  &$d(\mathcal L(\A),\mathcal L(\A^n))$ &$d(\mathcal L(\A^+)),\mathcal L(\A^n))$ &gain & deviation \tabularnewline
			\midrule 
			$[0.005, 0.025]$ & $37$ & $0.00800$ & $0.00291$ & $0.00382$ &\cellcolor{high} $1.31271$ &0.00010  \tabularnewline
			$[0.002,0.005]$ &$61$& $0.00384$ & $0.00144$ & $0.00177$ &\cellcolor{high} $1.22917$ &0.00033  \tabularnewline
			$[0.001,0.002]$ & $84$ & $0.00141$ & $0.00055$ & $0.00065$ &\cellcolor{high} $1.18182$ &0.00016  \tabularnewline
			$[0.0005,0.001]$ & $59$ & $0.00067$& $0.00026$ & $0.00033$&\cellcolor{high} $1.26923$ &0.00003  \tabularnewline	
			$[0.00005,0.0005]$ & $59$ & $0.00041$& $0.00016$ & $0.00019$&\cellcolor{high} $1.18750$&0.00002  \tabularnewline
			\bottomrule
		\end{tabular}}
	}

		\caption{Evaluation of the algorithm w.r.t. the elimination of pathological behaviours.}\label{tbl:summaryBenchmarkSubSuper}
	\end{table}

The experimental results are shown in Table~\ref{tbl:summaryBenchmarkSubSuper}, where each row represents the results of all pairs of $\A$ and $\A^+$ whose distance inside the range depicted in the first column. Interestingly, the language of the learned DFA is always closer to the one of $\A$ than the one of $\A^+$, and this for all distances
between $\A$ and $\A^+$. However as could be expected, since $\A^+$ is a DFA, the information gain remains close to 1. Hence, our experimentation can be seen as an evidence of capacity of the KV’s algorithm to partially eliminate  pathological behaviours.

\subsection{Words distribution }\label{subsection:wordDist}
We now discuss the impact of word distribution on the robustness of the KV's algorithm.
The parameter $\mu$ determines the average length of a random word ($\frac{1}{\mu}-1$).
Table~\ref{tbl:wordDistrabution} summarizes experimental results with different values of $\mu$ indicated on the first row. 
The other rows correspond to different values of the noise $p$ for $\A^{\rightarrow p}$. The cells (at the intersection of a pair ($p$, $\mu$)) contain the (average) information gain, where experiments have been done over twenty-two DFA. Note that the worst and best cases are always eliminated to avoid that the pathological cases perturb the average values. For values of $p$ that matter (i.e., when the gain is greater than 1), there is clear tendency for the gain to first increase w.r.t. $\mu$, reaching a maximum
about $\mu=0.01$ the value that we have chosen and then decrease. A possible explanation would be the following: too short words (i.e., big $\mu$) does not help
to discriminate between languages while  too long words (i.e., small $\mu$) lead to overfitting and does not reduce the noise.

\begin{table}[h]
	\centering
	\scalebox{.9}{
	{\renewcommand{\arraystretch}{1.5}
	\begin{tabular}{l*{5}{@{\hskip1em}r}}
		\toprule
		\backslashbox{$p$}{$\mu$} & $0.001$ & $0.005$ & $0.01$ & $0.05$ & $0.1$ \tabularnewline
		\midrule
		$0.01$  &\cellcolor{low}$0.059$ & \cellcolor{low}$0.067$ & \cellcolor{low}$0.078$ & \cellcolor{low}$0.184$ & \cellcolor{low}$0.317$ \tabularnewline
		$0.005$ & \cellcolor{low}$0.078$ & \cellcolor{low}$0.130$ & \cellcolor{low}$0.134$ & \cellcolor{low}$0.559$ & \cellcolor{low}$0.966$ \tabularnewline
		$0.0025$ & \cellcolor{low}$0.165$ & \cellcolor{low}$0.298$ & \cellcolor{low}$0.398$ &\cellcolor{medium}$1.246$ & \cellcolor{low}$0.823$ \tabularnewline
		$0.0015$ & \cellcolor{low}$0.465$ & \cellcolor{low}$0.671$ & \cellcolor{high}$2.267$ & \cellcolor{high}$2.074$ & \cellcolor{high}$1.651$ \tabularnewline
		$0.001$ & \cellcolor{high}$1.801$ & \cellcolor{high}$10.94$ & \cellcolor{high}$8.907$ & \cellcolor{high}$3.753$ & \cellcolor{high}$2.341$ \tabularnewline
		\bottomrule 
	\end{tabular}}
}
	\caption{Analysis of different distributions on $\Sigma^*$ }\label{tbl:wordDistrabution}
\end{table}

\begin{algorithm}[ht]
        \DontPrintSemicolon
        
        \KwIn{$L$, a language unknown to the algorithm}
        \KwIn{an integer $maxround$ ensuring termination}
        \SetKwFunction{AdaptedAngluin}{AdaptedAngluin}

	\BlankLine
	\KwData{integers $r$, $\mathit{rfinal}$ and $period$, a boolean $b$, and a data structure $Data$}
	\KwOut{a DFA}
	\BlankLine
	
	{\tt Initialize}($Data$)\;
	$r \leftarrow 0$\;
        $\mathit{rfinal} \leftarrow maxround$\;
	\BlankLine
	
	\tcp{The control of $maxround$ is unnecessary when $L$ is regular}
	\While{$r < maxround$}
	{
	 $\A_E \leftarrow $ {\tt Synthetize}($Data$)\;
	 
	 $(b,w) \leftarrow$ {\tt IsEquivalent}($\A_E$)
	  
	\lIf{$b$}\
        {\hspace*{0.6cm}$\mathit{rfinal} \leftarrow r$}\;
	{\hspace*{0.6cm}break}\;
	 {\tt Update}($Data,w$)\;
	 $r\leftarrow r+1$\;
      \lIf{$r\%period==0$}\
      {\hspace*{0.6cm}$\mathbf{TA}[r/period]\leftarrow \A_E$
      }
	}
    $i \leftarrow 1$\;
    $threshold \leftarrow c*{\bf Pr}_D(\Lan(\A_E))$

    \Repeat{$i*period > \mathit{rfinal}$}
	{
	  
	\lIf{$d(\Lan(TA[i]), \Lan(\A_E))\le threshold$}\
        {\hspace*{0.6cm}$\mathit{rfinal} \leftarrow r$}\;
        {\hspace*{0.6cm}return {\Minimize}($\mathbf{TA}[i]$)}\;
        \lElse
        {\hspace*{0.6cm}$i \leftarrow i+1$
      }
	}
	
	 \Return  {\Minimize}($\A_E$)

	\caption{Modified KV's algorithm for size reduction of the DFA}
	\label{algo:Adapted}    
\end{algorithm}
\subsection{Reduction of the size of the DFA}
Up to now the extracted DFA is obtained by using as exit condition that either the maximal number of rounds
is reached or the current learned automaton is declared equivalent. In practice, since the language of the noisy device is generally not regular, the current automaton of the algorithm is rarely declared equivalent and thus the final $\A_E$ is very often returned after running the maximal number of rounds. 
In other words, after the maximal number of rounds, chosen 250 here, the (minimal) extracted DFA has normally a larger size than the original one, on average about twice the original one. 
In view of eliminating over-fitting, we want  to reduce the size of the returned DFA. So we proceed as follows.
\begin{itemize}
    \item Every 10 rounds, we memorize the current automaton in an array $\mathbf{TA}$;
    \item When we exit the while loop, we compute the measure
    of $\Lan(\A_E)$, denoted $m={\bf Pr}_D(\Lan(\A_E))$ and define a threshold which is $c*m$, where we choose $c=10^{-3}$;
    \item Then we examine the saved DFAs by increasing size
    and select the first one whose distance from $\A_E$
    is smaller than the threshold and return it (after a minimization);
    \item Otherwise we return $\A_E$ (after a minimization).
\end{itemize}
This is formalized by Algorithm~\ref{algo:Adapted}.

Table~\ref{tbl:tradeoff} shows the experimental results in the following setting, where we randomly generated sixty DFA as before.
\begin{itemize}
    \item the size of the sixty DFA is randomly selected between 20 and 60;
    \item we only consider DFA with noisy outputs with the parameter $p$ between $10^{-2}$ and $10^{-3}$.
\end{itemize}
 In this table, $\widehat{\A}_E$ is the DFA returned by Algorithm~\ref{algo:Adapted} while $\A_E$ is the DFA returned by Algorithm~\ref{algo:PAC}. Similarly, $\widehat{gain}$ and
 $gain$ are the information gains corresponding to Algorithm~\ref{algo:Adapted} and Algorithm~\ref{algo:PAC}, respectively.
 We first observe that the information gains are very close
 whose better one can equally be $\widehat{gain}$ or
 $gain$. 
 As long as the information gain is less than one, the size reduction
 is not significant while as soon as information gain is more than one,
 the size decreases by about $\frac 1 3$. So we conclude that this adaptation is useful when the KV's algorithm performs well, i.e., robust against the introduced noise.

We also performed experimentations with a
threshold equal to $0.1m$, $0.01m$, and $0.005m$ not presented in this table. The results establish that this reduces more the size of the DFA but that the information gain is considerably worse than for  Algorithm~\ref{algo:PAC}
and most of the time less than one.

\begin{table}[h]
		\centering
		\scalebox{.9}{
		{\renewcommand{\arraystretch}{1.5}
		\begin{tabular}{l*{5}{r}}
			\toprule
			$ p $ & $|\widehat{\A}_E|$ & $|\A_E|$&$|\widehat{\A}_E|/|\A_E|$&$\widehat{gain}$ &$gain$ \tabularnewline
				\midrule
			$0.01$ & $182.480$ & $182.940$ &$0.99749$&\cellcolor{low} $0.11326$ &\cellcolor{low} $0.11425$  	\tabularnewline
			$0.005$ & $137.040$ & $141.700$ &$0.96711$ & \cellcolor{low} $0.50633$ & \cellcolor{low} $0.49839$ 	\tabularnewline
			$0.0025$ & $68.380$ & $94.500$ &$0.72360$ &\cellcolor{high} $2.64267$ &\cellcolor{high} $2.62994$  \tabularnewline
			$0.0015$ & $48.840$ & $77.480$ &$0.63036$ &\cellcolor{high} $3.33972$ &\cellcolor{high} $3.44385$  \tabularnewline
			$0.001$ & $44.600$ & $66.980$ &$0.66587$ &\cellcolor{high} $6.81623$  &\cellcolor{high} $6.49229$ \tabularnewline
			\bottomrule
		\end{tabular}}
	}
		
		\caption{Comparison between Algorithms~\ref{algo:PAC}
  and~\ref{algo:Adapted}.}\label{tbl:tradeoff}
	\end{table}

\section{Random languages versus structured languages}
\label{sec:theory}

Recall that in the precedent section, from the experimental results, we conjecture that KV’s algorithm is robust, when the noise is random, i.e., unstructured, and its quantity is small enough, such as for DFA with noisy output and with noisy input. This is however not the case for structured counter DFA, for which KV’s algorithm learns the noisy device itself instead of the original one whatever the quantity of noise.

In this section, we want to theoretically establish  that the main factor of the robustness
of the KV's  algorithm w.r.t.\ random noise is that almost surely randomness, in most cases, 
yields the perturbated language that is unstructured. We consider a language as structured if it can
be produced by some general device. Thus we identify the family of structured languages
with the family of recursively enumerable languages. More precisely, we show that almost surely DFA with noisy output leads to a language that is not recursively enumerable. We then demonstrate further that with a mild condition, almost surely DFA with noisy input yields also non-recursively enumerable language. As for the counter DFA, by definition, it is clearly recursively enumerable, thus not being studied further. 

The following lemma gives a simple means to establish that almost surely 
a random language is not recursively enumerable.

\begin{lem}
\label{lemma:nonre}
Let $R$ be a random language over $\Sigma$. Let $(w_n)_{n\in \nat}$ be a sequence of words of $\Sigma^*$.
Let $W_n=\{w_i\}_{i<n}$ and $\rho_n=\max_{W\subseteq W_n}{\bf Pr}(R \cap W_n=W)$.
Assume that $\lim_{n\rightarrow \infty} \rho_n=0$. 
Then, for all countable families of languages $\mathcal F$, almost surely $R \notin \mathcal F$.
In particular, almost surely $R$ is not a recursively enumerable language.
\end{lem}
\begin{proof}
Let us consider an arbitrary language $L$. 
Then, for all $n$,
$$ {\bf Pr}(R = L) \leq {\bf Pr}(R \cap W_n=L \cap W_n)\leq\rho_n.$$
Thus, ${\bf Pr}(R =L)=0$ and ${\bf Pr}(R \in \mathcal F)=\sum_{L\in \mathcal F}{\bf Pr}(R =L)=0$.
\end{proof}

From Lemma~\ref{lemma:nonre}, we immediately obtain that almost surely
the noisy output perturbation of any language is not recursively enumerable.
The proofs of the two next theorems use the same notations as those given in Lemma~\ref{lemma:nonre}.
\begin{thm}
\label{theorem:un}
	Let $L$ be a  language and $0<p<1$. Then almost surely $L^{\rightarrow p}$ is not a recursively enumerable language.
\end{thm}
\begin{proof}
Consider any enumeration  $(w_n)_{n\in \nat}$ of $\Sigma^*$ and any $W\subseteq W_n$.
The probability that $L^{\rightarrow p}\cap W_n$ is equal to $W$  is bounded
by $\max(p,1-p)^n$. Thus, $\rho_n\leq \max(p,1-p)^n$ and $\lim_{n\rightarrow \infty} \rho_n=0$.
\end{proof}

\begin{figure}[h]
	\centering
	
	\includegraphics[scale=0.87]{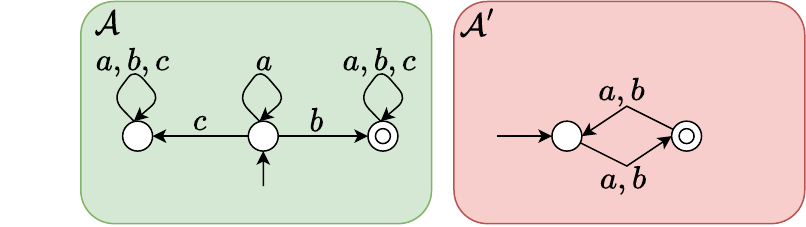}
	\caption{Two DFA}
	\label{fig:inOutProp2}
\end{figure}

We cannot get a similar result for the noisy input perturbation. Indeed consider the language
$\Sigma^*$, whatever the kind of noise brought to the input, the obtained language is still $\Sigma^*$.
With the kind of input noise that we study, consider the language that accepts words of odd length 
(see the automaton $\A'$ of Figure~\ref{fig:inOutProp2}).
Then the perturbed language with noisy input is unchanged, i.e., $\Lan(\A')=\Lan(\A'^{\leftarrow p})$.

However given a DFA $\mathcal A$, we can establish a mild condition
on $\mathcal A$ ensuring that almost surely the random language $\Lan(\mathcal A^{\leftarrow p})$ is not recursively enumerable. 
To this end, we now recall Markov chains with some important relative notions.

\noindent
{\bf Notation.}
Let $Q$ be a finite set of states.
Then $Dist(Q) = \{\Delta : Q \rightarrow \rat_{\geq 0} \mid \sum_{q\in Q}\Delta(q)=1\}$  is the set of \emph{rational distributions} over $Q$.

\pagebreak{}

\begin{defi}[Markov chain]
A \emph{finite Markov chain} is a tuple $\mathcal M=(Q, P)$  where:
\begin{itemize} \setlength\itemsep{0.5em}
	\item  $Q$ is a finite set of states;
	\item P is the transition function from $Q$ to $Dist(Q)$
    with $P(q)(q')$ also denoted $P(q,q')$.	
\end{itemize}
\end{defi}

\begin{defi}[Irreducibility and Periodicity]
Let $\mathcal M=(Q, P)$ be a finite Markov chain. Then:
\begin{itemize} \setlength\itemsep{0.5em}
    \item $\mathcal M$ is \emph{irreducible} if for any two states  
    $q, q'\in Q$, there exists $n \in \nat$ such that $P^n(q, q')>0$;
    \item Assume that $\mathcal M$ is irreducible and pick some $q_0\in Q$. Then the periodicicity of $\mathcal M$ denoted $period(\mathcal M)$ is defined by $period(\mathcal M)=\gcd(\{n>0 \mid P^n(q_0,q_0)>0\})$ (which is in fact independent of $q_0$).
\end{itemize}
\end{defi}

\noindent
{\bf Notation.}
As usual in the context of graphs, a bottom strongly connected component will be denoted by a BSCC.

We are now ready to present a mild condition on a DFA $\A$ such that almost surely 
the random language $\Lan(\mathcal A^{\leftarrow p})$ is not recursively enumerable. 

\begin{defi}[equal-length-distinguishing DFA]\label{def:eld}
Let $\A = (Q,F,\sigma,q_0)$ be a DFA.
We call $\A$ \emph{equal-length-distinguishing} if
there exist (possibly identical) BSCC $\mathcal C, \mathcal C'$ of $\A$,
$q_1\in \mathcal C \cap F$, $q'_1\in \mathcal C' \setminus F$, and $w,w'\in \Sigma^*$ such that
we have $q_1 = \sigma(q_0,w)$, $q'_1 = \sigma(q_0,w')$, and $|w|=|w'|$.
\end{defi}

\begin{thm}
\label{theorem:deux}
	Let $\Sigma$ be an alphabet with $|\Sigma|>1$.
	Let $\A = (Q,\sigma,q_0, F)$ be a DFA over $\Sigma$, $0<p<1$ and $\mathcal C, \mathcal C'$ some BSCC
	of $\A$ (possibly equal).
	Assume that $\A$ is equal-length-distinguishing.
	Then almost surely $\Lan(\A^{\leftarrow p})$ is not a recursively enumerable language.
\end{thm}
\begin{proof}
Let us denote  $\ell=|w|$ and let $a\in \Sigma$.
We build a Markov chain $\mathcal M$ from $\mathcal C$ as follows: every transition
	$q\xrightarrow{a}q'$ has probability $1-p$ and for all $b\neq a$, every transition $q\xrightarrow{b}q'$ has probability $\frac{p}{|\Sigma|-1}$.
	We proceed similarly from $\mathcal C'$ to build $\mathcal M'$. We denote $m=period(\mathcal M)$ and 
$m'=period(\mathcal M')$ in the following.

\noindent
Let us denote $\alpha_n$ (resp.  $\alpha'_n$ ) the probability in $\mathcal M$ (resp. $\mathcal M'$) that starting from $q_1$ 
	(resp.  $q'_1$), the current state at time $n$ is $q_1$ (resp. $q'_1$). For the sake of simplicity, we reuse $q_1$ and $q'_1$ as in Definition~\ref{def:eld}.  
        Since $\mathcal M$ and $\mathcal M'$ are irreducible with respectively periodicity $m$ and $m'$, $\lim_{n\rightarrow \infty} \alpha_{mn}$ (resp. $\lim_{n\rightarrow \infty} \alpha'_{m'n}$)
         exists and is positive. Let us denote $\alpha$ (resp. $\alpha'$) this limit. There exists $n_0$ such that for all $n\geq n_0$, 
         $\alpha_{mn}\geq \frac{\alpha}{2}$ and $ \alpha'_{m'n} \geq \frac{\alpha'}{2}$.
         
\noindent
Define $w_n=wa^{mm'(n+n_0)}$ for all $n\in \N$.        
The probability that $w_n$ is accepted by $\Lan(\A)^{\leftarrow p}$ is lower bounded by  the probability 
         that the prefix $w$ is unchanged (thus reaching $q_1$) and that after $mm'(n+n_0)$ steps the current state in $\mathcal M$ is $q_1$. Recall that 
         $q_1$ is a final state.
         So a lower bound is: $\min(p,1-p)^\ell \frac{\alpha}{2}$.

\noindent
The probability that $w_n$ is rejected by $\Lan(\A)^{\leftarrow p}$ is lower bounded by  the probability 
         that the prefix $w$ is changed into $w'$ (thus reaching $q'_1$) and that after $mm'(n+n_0)$ steps the current state in $\mathcal M'$ is $q'_1$. Note that 
         $q'_1$ is not a final state. 
         So a lower bound is: $\min(p,1-p)^\ell \frac{\alpha'}{2}$.
		
\noindent
Let $W\subseteq W_n$. The probability that $L^{\leftarrow p}\cap W_n$ is equal to $W$  is upper bounded
	by: $$\left(1-\min(p,1-p)^\ell \frac{\min(\alpha,\alpha')}{2}\right)^n$$ 
	
\noindent
Thus $\rho_n\leq \left(1-\min(p,1-p)^\ell \frac{\min(\alpha,\alpha')}{2}\right)^n$ and $\lim_{n\rightarrow \infty} \rho_n=0$.	
\end{proof}

The DFA $\A$ of Figure~\ref{fig:inOutProp2} that represents the formula `$a \text{ Until } b$' of temporal logic LTL
is equal-length-distinguishing. The corresponding pair of states consists of the accepting state and the leftmost one, both constituting a 
BSCC with only one of them being a final state. The DFA $\A'$ on the right part of Figure~\ref{fig:inOutProp2} is not equal-length-distinguishing. 
The whole DFA is a BSCC with only one final state. There does not exist a pair of $w, w'\in \Sigma^*$ satisfying the condition of Definition~\ref{def:eld}. 
Given a DFA $\A=(Q, \sigma, q_0, F)$, checking whether it is equal-length-distinguishing can be done in quadratic time with the following procedure.
\begin{enumerate}
    \item construct a new graph $G$ from $\A$  
    \begin{itemize}
        \item the set of vertices is $Q\times Q$;
        \item there is an edge $(q_1,q_2) \rightarrow (q'_1,q'_2)$ in $G$ if there are some transitions $q_1 \xrightarrow{a_1} q'_1$ and $q_2 \xrightarrow{a_2} q'_2$ in $\A$.
    \end{itemize}
    \item check whether there exists at least one vertex $(q_1,q_2)$ in some BSCC of $G$ with $q_1\in F$ and $q_2\notin F$, such that it is reachable from $(q_0,q_0)$ in $G$. The existence of such a vertex implies that $\A$ is equal-length-distinguishing. 
\end{enumerate}
The omitted correctness proof of this procedure is straightforward.

Since being equal-length-distinguishing is a sufficient condition for ensuring that almost surely $\Lan(\A^{\leftarrow p})$ is not a recursively enumerable language, we want to investigate whether it is necessary.
The next proposition shows a particular case when it is the case.

\begin{prop}\label{prop1}
	Let $\Sigma$ be an alphabet with $|\Sigma|>1$.
	Let $\A = (Q,\sigma,q_0, F)$ be a DFA that is not equal-length-distinguishing and such that every circuit of $\A$ belongs to a BSCC.
	Then, for every sampling $L'$ of $\Lan(\A^{\leftarrow p})$, $L'$ is regular.
\end{prop}
\begin{proof}
Pick some $n_0\in \N$ such that for all $w$ with $|w|\geq n_0$ and $q_0\xrightarrow{w} q$ implies that $q$ belongs to some BSCC.
Observe now that, since $\A$ is not equal-length-distinguishing, for  words $w,w'$ with $|w|=|w'|\geq n_0$,
$w\in L$ iff $w'\in L$. Thus, for every sampling $L'$ of $\Lan(\A^{\leftarrow p})$, $L'=(L' \cap \Sigma^{<n_0}) \cup (L\cap \Sigma^{\geq n_0})$
implying that $L'$ is regular.
\end{proof}
\begin{figure}[h]
\begin{footnotesize}
\begin{center}
\begin{tikzpicture}[xscale=0.9,yscale=0.99]
\begin{scope}[auto,every node/.style={draw,circle,minimum size=20pt}]

\path (3,0) node [draw,circle,inner sep=2pt](l3) {$q_0$};
\path (9,0) node [draw,circle,double,inner sep=2pt](lf) {$q_f$};
\path (6,0) node [draw,circle,inner sep=2pt](lr) {$q_r$};

\end{scope}
 
\draw[arrows=-latex'] (2,0) -- (l3);
\draw[arrows=-latex'] (l3) --(6,1.2)--(lf)node[pos=0,above] {$a$};
\draw[arrows=-latex'] (lf) --(6,-1.5)--(l3)node[pos=0,below] {$b$};
\draw[-latex'] (lf) .. controls +(-300:40pt) and +(-240:40pt) .. (lf)
node[pos=0.5,above] {$a$};
\draw[-latex'] (l3) .. controls +(-300:40pt) and +(-240:40pt) .. (l3)
node[pos=0.5,above] {$b$};
\draw[arrows=-latex'] (l3) --(lr)node[pos=0.5,above] {$c$};
\draw[arrows=-latex'] (lf) -- (lr)node[pos=0.5,above] {$c$};
\draw[-latex'] (lr) .. controls +(300:40pt) and +(240:40pt) .. (lr)
node[pos=0.5,above] {$\Sigma$};

\end{tikzpicture}
\end{center}
\end{footnotesize}
\caption{A DFA $\A$ with $\Lan(\A)=(a+b)^*a$}
\label{fig:counter-example}
\end{figure}

However  in general this condition is not necessary.
Observe that we establish the next proposition using a generalization of Lemma~\ref{lemma:nonre}.
\begin{prop}\label{prop2}
Let $\A$ be the DFA of Figure~\ref{fig:counter-example}.
Then, $\A$ is not equal-length-distinguishing while almost surely $\Lan(\A^{\leftarrow \frac 2 3})$ is not recursively enumerable.
\end{prop}
\begin{proof}
There is a single BSCC  with a single state $\{q_r\}$. So $\A$ is not equal-length-distinguishing.
Let $w\neq \lambda$ be a word  with $|w|=n$ and denote $\tilde{w}$ the random word obtained by the noisy perturbation. 
Observe that every letter of $\tilde{w}$ is uniformly distributed over $\Sigma$.
So the probability that $\tilde{w}$ does not contain a $c$ is $(\frac 2 3)^n$ and the conditional probability that $\tilde{w}$
belongs to $\Lan(\A^{\leftarrow \frac 2 3})$ knowing that it does not contain a $c$ is $\frac 1 2$.

\noindent
Fix some $0<\rho<1$. The probability that for all words $w\in \Sigma^n$, $\tilde{w}$ 
contains a $c$ is equal to $(1-(\frac 2 3)^n)^{3^n}\leq e^{-2^n}$. Pick an increasing sequence $(n_k)_{k\in \N}$ such $\sum_{k\in \N} e^{-2^{n_k}}\leq 1-\rho$.
Then with probability at least $\rho$, for all $k$, there is a word  $w_k\in \Sigma^{n_k}$ such that $\tilde{w}_k$ does not contain a $c$. Letting $\rho$
go to 1, almost surely there is an infinite number of words $w$ such that  $\tilde{w}\in (a+b)^+$. 

\noindent
Let us consider an arbitrary language $L'$ and $(w_n)_{n\in \N}$ be an enumeration of $\Sigma^+$.
Then almost surely there is an infinite number of $w_n$ such that $\tilde{w}_n$ belong to $(a+b)^+$. 
Recall that for such a word, the probability that it belongs to $\Lan(\A^{\leftarrow \frac 2 3})$ is equal to $\frac 1 2$.
Let $W_n$ be the \emph{random} set of the first $n^{th}$ such words.
Then for all $n$, $ {\bf Pr}(L' =\Lan(\A^{\leftarrow \frac 2 3})) \leq {\bf Pr}(L' \cap W_n=\Lan(\A^{\leftarrow \frac 2 3}) \cap W_n)=2^{-n}$.

Thus ${\bf Pr}(L' =\Lan(\A^{\leftarrow \frac 2 3}))=0$ and ${\bf Pr}(\Lan(\A^{\leftarrow \frac 2 3}) \in \mathcal F)=\sum_{L'\in \mathcal F}{\bf Pr}(L' =\Lan(\A^{\leftarrow \frac 2 3}))=0$
for $ \mathcal F$ a countable family of languages.
\end{proof}

To show the soundness of the structural criterion described in Theorem~\ref{theorem:deux} with experiments and comparisons, we have refined our experiments on DFA with noisy inputs by partitioning the randomly generated DFA, depending on whether they are equal-length-distinguishing or not.  

We have chosen $|\Sigma| = 3$ since with greater size, it was difficult to generate DFAs that do not satisfy the hypotheses.
Tables~\ref{tbl:noisyWoSpecialProperty} and~\ref{tbl:noisySpecialProperty} summarize these experiments.
The last rows of the tables (where the information gain is greater than one) confirm our conjecture. More precisely, for any equal-length-distinguishing DFA $\A$, as almost surely $\Lan(\A^{\leftarrow p})$ is not a recursively enumerable language, then beyond a certain threshold, the robustness of the algorithm increases quickly which is not the case in Table~\ref{tbl:noisySpecialProperty}.

\begin{table}[h]
	\centering
	\small
	\scalebox{.9}{
	{\renewcommand{\arraystretch}{1.5}
	\begin{tabular}{l*{6}{@{\hskip1em}r}}
		\toprule
		Range& \# & $d(\mathcal L(\A),\mathcal L(\A^{\leftarrow p}))$  & $d(\mathcal L(\A),\mathcal L(\A_E))$ &  $d(\mathcal L(\A^{\leftarrow p})),\mathcal L(\A_E))$ &gain & deviation \tabularnewline
		\midrule 
		$[0.005, 0.025]$ & $85$ & $0.01114$ & $0.03604$ & $0.04345$ &\cellcolor{low} $0.30902$&0.05162	\tabularnewline
		$[0.002, 0.005]$ & $81$ & $0.00338$ & $0.00421$ & $0.00747$ &\cellcolor{low} $0.80443$&0.02793 \tabularnewline
		$[0.001, 0.002]$ & $25$ & $0.00142$ & $0.00035$ & $0.00174$ &\cellcolor{high} $4.09784$&0.00062 \tabularnewline
		$[0.0005, 0.001]$ & $16$ & $0.00071$ & $0.00006$ & $0.00077$ &\cellcolor{high} $11.08439$&0.00006
\tabularnewline	
		\bottomrule
	\end{tabular}}
       }
	\caption{Experiments on equal-length-distinguishing DFA}
	\label{tbl:noisyWoSpecialProperty}
\end{table}

\begin{table}[h]
	\centering
	\small
	\scalebox{.9}{
	{\renewcommand{\arraystretch}{1.5}
	\begin{tabular}{l*{6}{@{\hskip1em}r}}
		\hline
		Range& \# & $d(\mathcal L(\A),\mathcal L(\A^{\leftarrow p}))$  & $d(\mathcal L(\A),\mathcal L(\A_E))$ &  $d(\mathcal L(\A^{\leftarrow p}),\mathcal L(\A_E))$ &gain&deviation \tabularnewline
		\midrule
		$[0.005, 0.025]$ & $36$ & $0.01089$ & $0.02598$ & $0.03410$ &\cellcolor{low} $0.41905$&0.06152 \tabularnewline
		$[0.002, 0.005]$ & $49$ & $0.00308$ & $0.00387$ & $0.00646$ &\cellcolor{low} $0.79628$&0.03763 \tabularnewline
		$[0.001, 0.002]$ & $35$ & $0.00136$ & $0.00057$ & $0.00182$ &\cellcolor{high} $2.39863$&0.00072 \tabularnewline
		$[0.0005, 0.001]$ & $36$ & $0.00075$ & $0.00063$ & $0.00130$ &\cellcolor{medium} $1.18583$&0.00005 \tabularnewline
		\bottomrule
	\end{tabular}}
}
	\caption{Experiments on non equal-length-distinguishing DFA}
	\label{tbl:noisySpecialProperty}
\end{table}

\section{Conclusion}
\label{sec:con}

We have studied how the PAC-version of KV's algorithm behaves for devices which are obtained
from a DFA by introducing noise. More precisely, we have investigated whether KV's algorithm reduces the noise 
producing a DFA closer to the original one than the noisy device.
We have  considered four kinds of noise either being random  or structured. We have shown that, on average, KV's algorithm
behaves well for random noise but not for structured one. We have completed our study by establishing 
that almost surely the random noisy devices produce a non recursively enumerable language
confirming the relevance of the structural criterion for robustness of KV's algorithm. 

There are several directions for future work.
In the short run, we want investigate whether our results are specific to the KV’s algorithm
or valid for all  variants of the Angluin’s algorithm.
In another direction, KV's algorithm has no information about the original DFA. It would be interesting to introduce a priori knowledge
and design more efficient algorithms. For instance, the algorithm could take as input the maximal size of the original DFA or 
a regular language that is a superset of the original language.
In our setting the noise resulted in a noisy device which, once obtained, answers membership queries deterministically. A  different form of noise to be studied would be that the answer to a query is randomly noisy meaning that for the same repeated query, 
different answers could occur. 

 Finally the language inference capacity of recurrent neural networks (RNN) especially on DFA has been  demonstrated by recent work~\cite{weiss18,MayrY18}. 
So an interesting subsequent work would be to examine whether the KV's algorithm 
reduces the noise introduced by such RNNs.

\bibliographystyle{alphaurl}
\bibliography{bib}

\newcommand{\etalchar}[1]{$^{#1}$}
\begin{thebibliography}{ACSvdB20}

\bibitem[ACSvdB20]{Kousar20}
Kousar Aslam, Loek Cleophas, Ramon Schiffelers, and Mark van~den Brand.
\newblock Interface protocol inference to aid understanding legacy software
  components.
\newblock {\em Softw. Syst. Model.}, 19(6):1519–1540, nov 2020.
\newblock \href {https://doi.org/10.1007/s10270-020-00809-2}
  {\path{doi:10.1007/s10270-020-00809-2}}.

\bibitem[AL87]{AngluinL87}
Dana Angluin and Philip~D. Laird.
\newblock Learning from noisy examples.
\newblock {\em Mach. Learn.}, 2(4):343--370, 1987.
\newblock \href {https://doi.org/10.1023/A:1022873112823}
  {\path{doi:10.1023/A:1022873112823}}.

\bibitem[Ang87]{Angluin87}
Dana Angluin.
\newblock Learning regular sets from queries and counterexamples.
\newblock {\em Inf. Comput.}, 75(2):87--106, 1987.
\newblock \href {https://doi.org/10.1016/0890-5401(87)90052-6}
  {\path{doi:10.1016/0890-5401(87)90052-6}}.

\bibitem[BEK02]{BSHOUTY02}
Nader~H. Bshouty, Nadav Eiron, and Eyal Kushilevitz.
\newblock {PAC learning with nasty noise}.
\newblock {\em Theoretical Computer Science}, 288(2):255--275, 2002.
\newblock Algorithmic Learning Theory.
\newblock \href {https://doi.org/10.1016/S0304-3975(01)00403-0}
  {\path{doi:10.1016/S0304-3975(01)00403-0}}.

\bibitem[BF72a]{Biermann72b}
A.~W. Biermann and J.~A. Feldman.
\newblock On the synthesis of finite-state machines from samples of their
  behavior.
\newblock {\em IEEE Trans. Comput.}, 21(6):592–597, jun 1972.
\newblock \href {https://doi.org/10.1109/TC.1972.5009015}
  {\path{doi:10.1109/TC.1972.5009015}}.

\bibitem[BF72b]{Biermann72}
Alan~W. Biermann and Jerome~A. Feldman.
\newblock A survey of results in grammatical inference.
\newblock In S.~Watanabe, editor, {\em Frontiers of Pattern Recognition}, pages
  31--54. Academic Press, New York, 1972.
\newblock \href {https://doi.org/10.1016/B978-0-12-737140-5.50007-5}
  {\path{doi:10.1016/B978-0-12-737140-5.50007-5}}.

\bibitem[BHKL09]{Bollig09}
Benedikt Bollig, Peter Habermehl, Carsten Kern, and Martin Leucker.
\newblock Angluin-style learning of nfa.
\newblock In {\em Proceedings of the 21st International Joint Conference on
  Artificial Intelligence}, IJCAI'09, page 1004–1009. Morgan Kaufmann
  Publishers Inc., 2009.
\newblock \href {https://doi.org/10.5555/1661445.1661605}
  {\path{doi:10.5555/1661445.1661605}}.

\bibitem[CE07]{Clark07}
Alexander Clark and R\'{e}mi Eyraud.
\newblock Polynomial identification in the limit of substitutable context-free
  languages.
\newblock {\em Journal of Machine Learning Research}, 8:1725–1745, dec 2007.
\newblock \href {https://doi.org/10.5555/1314498.1314556}
  {\path{doi:10.5555/1314498.1314556}}.

\bibitem[CL10]{Cassandras10}
Christos~G. Cassandras and Stephane Lafortune.
\newblock {\em Introduction to Discrete Event Systems}.
\newblock Springer Publishing Company, Incorporated, 2010.
\newblock \href {https://doi.org/10.1007/978-0-387-68612-7}
  {\path{doi:10.1007/978-0-387-68612-7}}.

\bibitem[Cla10]{Clark2010}
Alexander Clark.
\newblock Distributional learning of some context-free languages with a
  minimally adequate teacher.
\newblock In {\em Grammatical Inference: Theoretical Results and Applications},
  ICGI'10, page 24–37. Springer-Verlag, 2010.
\newblock \href {https://doi.org/10.5555/1886263.1886269}
  {\path{doi:10.5555/1886263.1886269}}.

\bibitem[Dec97]{Decatur97}
Scott~E. Decatur.
\newblock Pac learning with constant-partition classification noise and
  applications to decision tree induction.
\newblock In {\em Proceedings of the Fourteenth International Conference on
  Machine Learning}, ICML '97, page 83–91, San Francisco, CA, USA, 1997.
  Morgan Kaufmann Publishers Inc.

\bibitem[FBLJ{\etalchar{+}}21]{Furelos21}
Daniel Furelos-Blanco, Mark Law, Anders Jonsson, Krysia Broda, and Alessandra
  Russo.
\newblock Induction and exploitation of subgoal automata for reinforcement
  learning.
\newblock {\em J. Artif. Int. Res.}, 70:1031–1116, may 2021.
\newblock \href {https://doi.org/10.1613/jair.1.12372}
  {\path{doi:10.1613/jair.1.12372}}.

\bibitem[FH17]{Brostean17b}
Paul Fiterau{-}Brostean and Falk Howar.
\newblock Learning-based testing the sliding window behavior of {TCP}
  implementations.
\newblock In Laure Petrucci, Cristina Seceleanu, and Ana Cavalcanti, editors,
  {\em Critical Systems: Formal Methods and Automated Verification -
  FMICS-AVoCS 2017, Turin, Italy, September 18-20, 2017, Proceedings}, volume
  10471 of {\em Lecture Notes in Computer Science}, pages 185--200. Springer,
  2017.
\newblock \href {https://doi.org/10.1007/978-3-319-67113-0_12}
  {\path{doi:10.1007/978-3-319-67113-0_12}}.

\bibitem[FJV16]{Brostean16}
Paul Fiterau{-}Brostean, Ramon Janssen, and Frits~W. Vaandrager.
\newblock Combining model learning and model checking to analyze {TCP}
  implementations.
\newblock In Swarat Chaudhuri and Azadeh Farzan, editors, {\em 28th
  International Conference on Computer Aided Verification, {CAV} 2016, Toronto,
  ON, Canada, July 17-23, 2016, Part {II}}, volume 9780 of {\em Lecture Notes
  in Computer Science}, pages 454--471. Springer, 2016.
\newblock \href {https://doi.org/10.1007/978-3-319-41540-6_25}
  {\path{doi:10.1007/978-3-319-41540-6_25}}.

\bibitem[FLP{\etalchar{+}}17]{Brostean17}
Paul Fiterau{-}Brostean, Toon Lenaerts, Erik Poll, Joeri de~Ruiter, Frits~W.
  Vaandrager, and Patrick Verleg.
\newblock Model learning and model checking of {SSH} implementations.
\newblock In Hakan Erdogmus and Klaus Havelund, editors, {\em Proceedings of
  the 24th {ACM} {SIGSOFT} International {SPIN} Symposium on Model Checking of
  Software, Santa Barbara, CA, USA, July 10-14, 2017}, pages 142--151. {ACM},
  2017.
\newblock \href {https://doi.org/10.1145/3092282.3092289}
  {\path{doi:10.1145/3092282.3092289}}.

\bibitem[Gol78]{Gold1978}
E~Mark Gold.
\newblock Complexity of automaton identification from given data.
\newblock {\em Information and Control}, 37(3):302 -- 320, 1978.
\newblock \href {https://doi.org/10.1016/S0019-9958(78)90562-4}
  {\path{doi:10.1016/S0019-9958(78)90562-4}}.

\bibitem[HIS{\etalchar{+}}12]{HowarISBJ12}
Falk Howar, Malte Isberner, Bernhard Steffen, Oliver Bauer, and Bengt Jonsson.
\newblock Inferring semantic interfaces of data structures.
\newblock In Tiziana Margaria and Bernhard Steffen, editors, {\em ISoLA 2012,
  Heraklion, Crete, Greece, October 15-18, 2012, Part {I}}, volume 7609 of {\em
  Lecture Notes in Computer Science}, pages 554--571. Springer, 2012.
\newblock \href {https://doi.org/10.1007/978-3-642-34026-0_41}
  {\path{doi:10.1007/978-3-642-34026-0_41}}.

\bibitem[Hoe63]{Hoff63}
Wassily Hoeffding.
\newblock Probability inequalities for sums of bounded random variables.
\newblock {\em Journal of the American Statistical Association},
  58(301):13--30, 1963.
\newblock \href {https://doi.org/10.2307/2282952} {\path{doi:10.2307/2282952}}.

\bibitem[JMKO20]{Natasha20}
Natasha~Yogananda Jeppu, Thomas Melham, Daniel Kroening, and John O'Leary.
\newblock Learning concise models from long execution traces.
\newblock In {\em Proceedings of the 57th ACM/EDAC/IEEE Design Automation
  Conference}, DAC '20. IEEE Press, 2020.
\newblock \href {https://doi.org/10.5555/3437539.3437631}
  {\path{doi:10.5555/3437539.3437631}}.

\bibitem[Kea98]{Kearns98}
Michael~J. Kearns.
\newblock Efficient noise-tolerant learning from statistical queries.
\newblock {\em J. {ACM}}, 45(6):983--1006, 1998.
\newblock \href {https://doi.org/10.1145/293347.293351}
  {\path{doi:10.1145/293347.293351}}.

\bibitem[KV94]{Kearns94}
Michael~J. Kearns and Umesh~V. Vazirani.
\newblock {\em An Introduction to Computational Learning Theory}.
\newblock {MIT} Press, 1994.
\newblock \href {https://doi.org/10.7551/mitpress/3897.001.0001}
  {\path{doi:10.7551/mitpress/3897.001.0001}}.

\bibitem[MP91]{Maler91}
Oded Maler and Amir Pnueli.
\newblock On the learnability of infinitary regular sets.
\newblock In {\em Proceedings of the Fourth Annual Workshop on Computational
  Learning Theory}, COLT '91, page 128–138. Morgan Kaufmann Publishers Inc.,
  1991.
\newblock \href {https://doi.org/10.5555/114836.114848}
  {\path{doi:10.5555/114836.114848}}.

\bibitem[MY18]{MayrY18}
Franz Mayr and Sergio Yovine.
\newblock Regular inference on artificial neural networks.
\newblock In Andreas Holzinger, Peter Kieseberg, A~Min Tjoa, and Edgar~R.
  Weippl, editors, {\em Proceedings of International Cross-Domain Conference
  for Machine Learning and Knowledge Extraction, {CD-MAKE} 2018, Hamburg,
  Germany, August 27-30, 2018}, volume 11015 of {\em Lecture Notes in Computer
  Science}, pages 350--369. Springer, 2018.
\newblock \href {https://doi.org/10.1007/978-3-319-99740-7_25}
  {\path{doi:10.1007/978-3-319-99740-7_25}}.

\bibitem[Qui86]{Quinlan86}
J.~R. Quinlan.
\newblock The effect of noise on concept learning.
\newblock In {\em Machine Learning, An Artificial Intelligence Approach Volume
  II}, chapter~6, pages 149--166. Morgan Kaufmann, 1986.

\bibitem[RS89]{Rivest89}
R.~L. Rivest and R.~E. Schapire.
\newblock Inference of finite automata using homing sequences.
\newblock In {\em Proceedings of the twenty-first annual ACM symposium on
  Theory of computing}, STOC '89, page 411–420. Association for Computing
  Machinery, 1989.
\newblock \href {https://doi.org/10.1145/73007.73047}
  {\path{doi:10.1145/73007.73047}}.

\bibitem[SG09]{Shahbaz09}
Muzammil Shahbaz and Roland Groz.
\newblock Inferring mealy machines.
\newblock In {\em Proceedings of the 2nd World Congress on Formal Methods}, FM
  '09, page 207–222. Springer-Verlag, 2009.
\newblock \href {https://doi.org/10.1007/978-3-642-05089-3_14}
  {\path{doi:10.1007/978-3-642-05089-3_14}}.

\bibitem[Slo95]{Sloan95}
Robert~H. Sloan.
\newblock Four types of noise in data for pac learning.
\newblock {\em Inf. Process. Lett.}, 54(3):157–162, may 1995.
\newblock \href {https://doi.org/10.1016/0020-0190(95)00016-6}
  {\path{doi:10.1016/0020-0190(95)00016-6}}.

\bibitem[Sol64]{Solomonoff64}
Ray~J. Solomonoff.
\newblock A formal theory of inductive inference.
\newblock {\em Inf. Control.}, 7(1, 2):1--22, 224--254, 1964.
\newblock \href {https://doi.org/10.1016/S0019-9958(64)90223-2}
  {\path{doi:10.1016/S0019-9958(64)90223-2}}.

\bibitem[Val84]{Valiant84}
Leslie~G. Valiant.
\newblock A theory of the learnable.
\newblock {\em Commun. {ACM}}, 27(11):1134--1142, 1984.
\newblock \href {https://doi.org/10.1145/1968.1972}
  {\path{doi:10.1145/1968.1972}}.

\bibitem[vdA12]{vanderalst12}
Wil M.~P. van~der Aalst.
\newblock Process mining.
\newblock {\em CACM}, 55(8):76--83, 2012.
\newblock \href {https://doi.org/10.1145/2240236.2240257}
  {\path{doi:10.1145/2240236.2240257}}.

\bibitem[WGY18]{weiss18}
Gail Weiss, Yoav Goldberg, and Eran Yahav.
\newblock Extracting automata from recurrent neural networks using queries and
  counterexamples.
\newblock In Jennifer Dy and Andreas Krause, editors, {\em Proceedings of the
  35th International Conference on Machine Learning}, volume~80, pages
  5247--5256. PMLR, 10--15 Jul 2018.
\newblock \href {https://doi.org/10.48550/arXiv.1711.09576}
  {\path{doi:10.48550/arXiv.1711.09576}}.

\bibitem[Wha74]{Wharton1974}
R.~M. Wharton.
\newblock Approximate language identification.
\newblock {\em Information and Control}, 26(3):236 -- 255, 1974.
\newblock \href {https://doi.org/10.1016/S0019-9958(74)91369-2}
  {\path{doi:10.1016/S0019-9958(74)91369-2}}.

\bibitem[YC10]{Yoshinaka2010}
Ryo Yoshinaka and Alexander Clark.
\newblock Polynomial time learning of some multiple context-free languages with
  a minimally adequate teacher.
\newblock In {\em Proceedings of the 15th and 16th International Conference on
  Formal Grammar}, FG'10/FG'11, page 192–207, Berlin, Heidelberg, 2010.
  Springer-Verlag.
\newblock \href {https://doi.org/10.1007/978-3-642-32024-8_13}
  {\path{doi:10.1007/978-3-642-32024-8_13}}.

\end{thebibliography}

\end{document}